\documentclass[5p]{elsarticle}

\usepackage{graphicx}
\usepackage{amsmath}
\usepackage{amsfonts}

\usepackage{color}

\usepackage{natbib}

\journal{Polymer}

\begin{document}

\begin{frontmatter}



\title{Disentanglement Effects on Welding Behaviour of Polymer Melts during the Fused-Filament-Fabrication Method for Additive Manufacturing}

\author{C. McIlroy}

\author{P.D. Olmsted}

\address{Department of Physics, and Institute for Soft Matter Synthesis and Metrology, Georgetown University, Washington DC, 20057, USA}

\begin{abstract}

Although 3D printing has the potential to transform manufacturing processes, the strength of printed parts often does not rival that of traditionally-manufactured parts. The fused-filament fabrication method involves melting a thermoplastic, followed by layer-by-layer extrusion of the molten viscoelastic material to fabricate a three-dimensional object. The strength of the welds between layers is controlled by interdiffusion and entanglement of the melt across the interface. However, diffusion slows down as the printed layer cools towards the glass transition temperature. Diffusion is also affected by high shear rates in the nozzle, which significantly deform and disentangle the polymer microstructure prior to welding. In this paper, we model non-isothermal polymer relaxation, entanglement recovery, and diffusion processes that occur post-extrusion to investigate the effects that typical printing conditions and amorphous (non-crystalline) polymer rheology have on the ultimate weld structure. Although we find the weld thickness to be of the order of the polymer size, the structure of the weld is anisotropic and relatively disentangled; reduced mechanical strength at the weld is attributed to this lower degree of entanglement.

\end{abstract}

\begin{keyword}

Fused filament fabrication, polymer melt, welding, disentanglement, non-isothermal

\end{keyword}

\end{frontmatter}

\raggedbottom

\section{Introduction}

Fused filament fabrication (FFF) \cite{Chua:2003} has become an essential tool for the rapid fabrication of custom parts via additive manufacturing. Although there are numerous advantages to this technique \cite{Turner:2014}, including ease of use, cost and flexibility, improving the strength of printed parts to rival that of traditionally-manufactured parts remains an underlying issue. 

\begin{figure*}[t]
\centerline{\includegraphics[width=16cm]{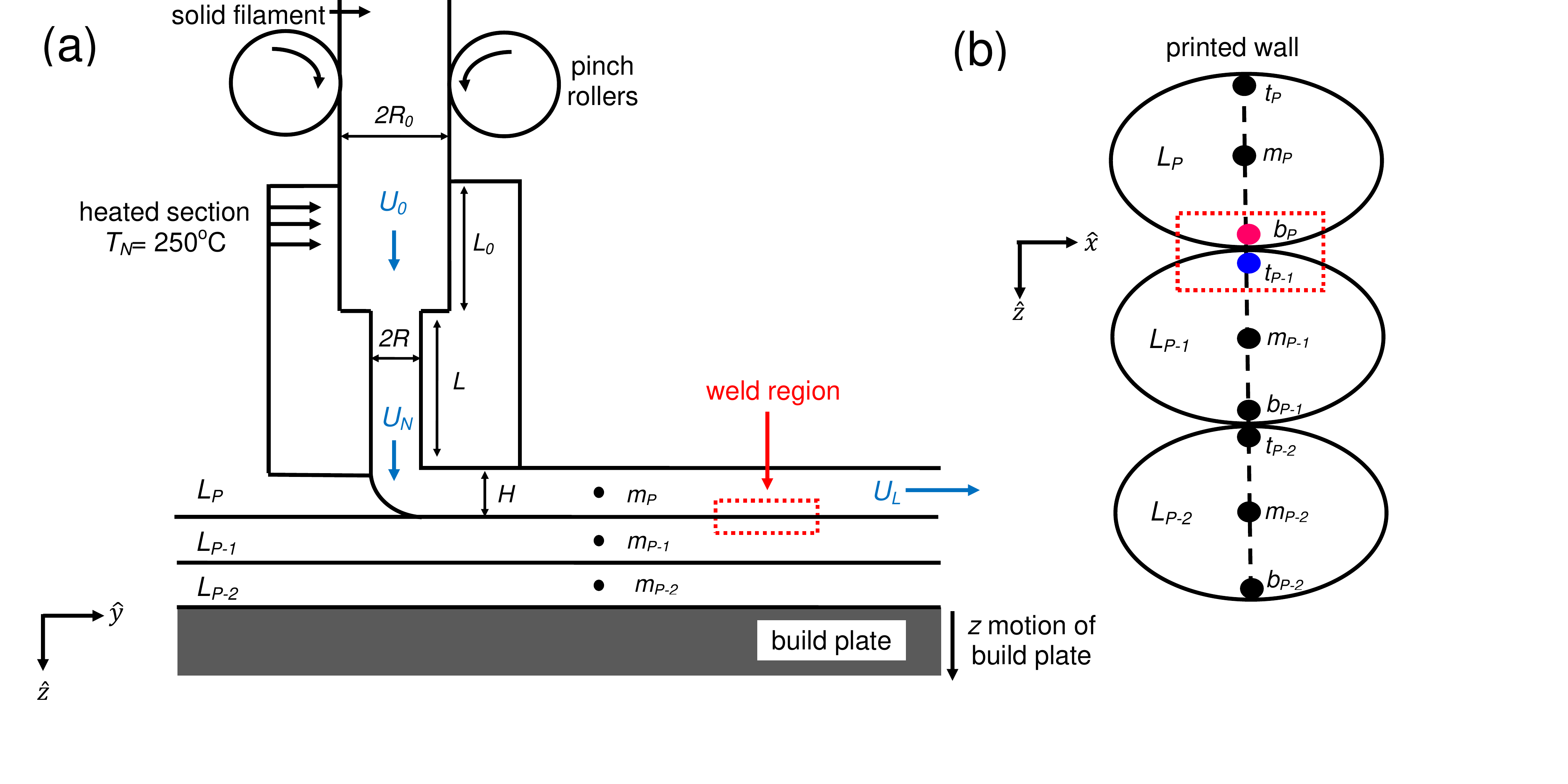}}
\caption{Simple schematic of typical FFF process, as described in text. (a) In the frame of the moving nozzle, the melt exits the nozzle at speed $U_N$ and the build plate moves at speed $U_L$ in the $\hat{y}$-direction. The current printed layer is denoted $\mathtt{L_{p}}$ and the middle of the layer is denoted $\mathtt{m_{p}}$. (b) Printed wall geometry with sites at the top $\mathtt{t}$, bottom $\mathtt{b}$ and middle $\mathtt{m}$ of each layer labelled. Welds occur at the interface between layers. See Appendix, Tables \ref{tab:polycarbonate}, \ref{tab:speeds} for typical model parameters.}
\label{fig:process}
\end{figure*}

The most common printing materials are amorphous polymer melts such as linear polycarbonate (PC) \cite{Hill:2014} and acrylonitrile butadiene styrene (ABS) \cite{Turner:2014}, a melt containing rubber nano-particles that provide toughness even at low temperatures. FFF printers can also handle semi-crystalline polymers such as poly-lactic acid (PLA) \cite{Drummer:2012}. The printing process involves melting a solid filament of the printing material and extruding it through a nozzle. To fabricate a three-dimensional object the melt is deposited layer-by-layer, as illustrated in Fig. \ref{fig:process}. The key to ensuring the strength of the final printed part is successful interdiffusion and re-entanglement of the polymer melt across the layer-layer interfaces.

In general, the weld strength of a polymer-polymer interface grows sub-diffusively with welding time as $t^{1/4}$ until the bulk strength plateau is reached \cite{Jud:1981, Kline:1988, Schnell:1999}. Several molecular mechanisms are proposed to explain this scaling. Since the weld thickness arising from interpenetration depth also scales as $t^{1/4}$ until the radius of gyration $R_g$ is reached due to polymer reptation, one suggested mechanism is that this interpenetration depth determines the weld strength \cite{Wool:1981, Schnell:1999, Brown:2001}. Others suggest that the formation of bridges between the surfaces is the key strengthening mechanism \cite{deGennes:1981, Prager:1981}. Both approaches are motivated by the idea of entanglement formation across the interface \cite{Robbins:2013}. Whilst some studies assume a simple proportionality between the interpenetration distance and entanglement formation \cite{Wool:1995, Wool:1981, Prager:1981, Benkoski:2002}, others assume a minimum interpenetration distance for an entanglement to form \cite{Brown:2001, Adolf:1985}. Few experiments find that diffusion by the radius of gyration $R_g$ is required to achieve bulk strength \cite{Schach:2008}, but in many cases welds reach bulk strength in much shorter times \cite{Wool:1995}.

\begin{figure*}[t!]
\centerline{\includegraphics[width=12cm]{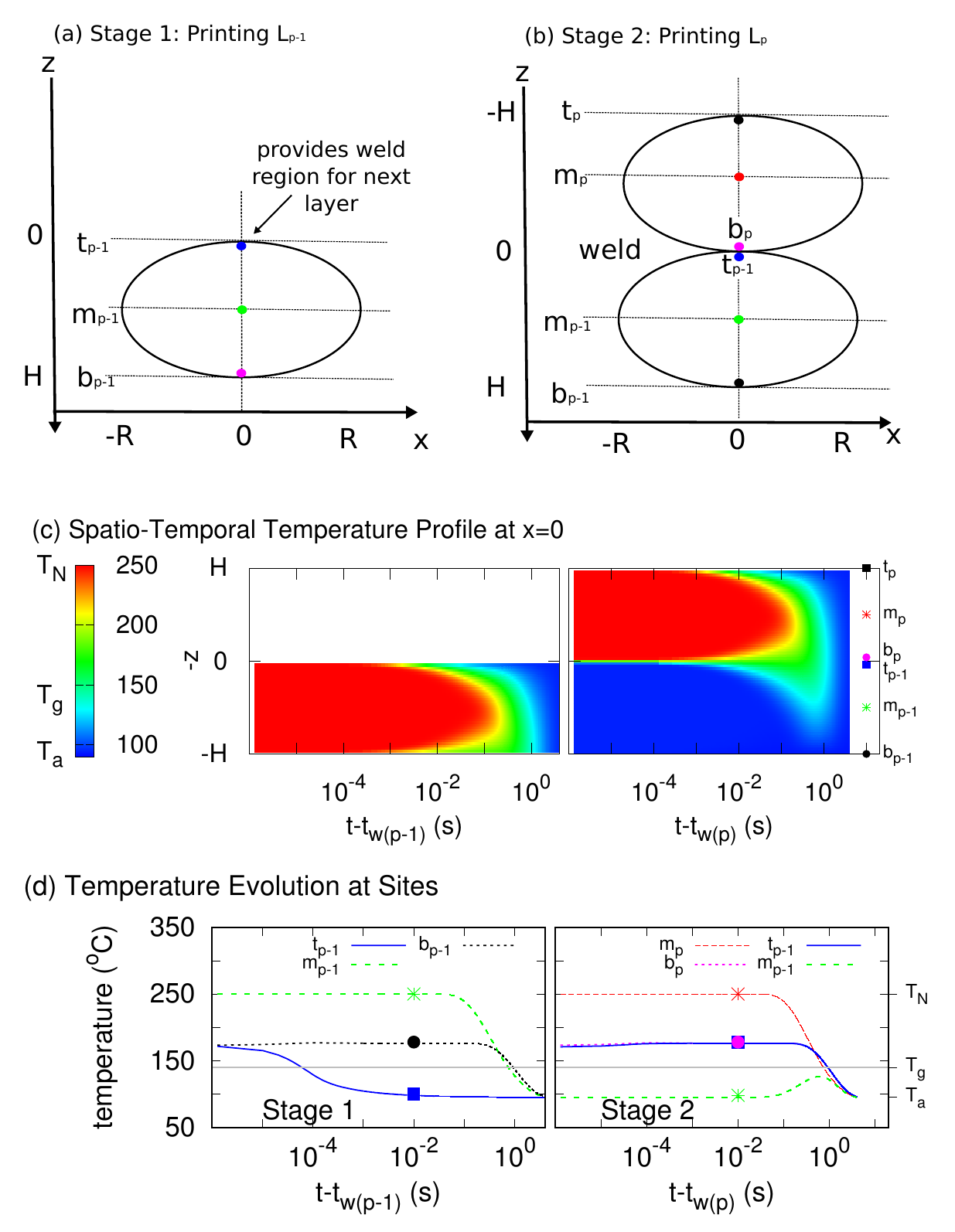}}
\caption{Schematic illustrating (a) first printing stage where $\text{L}_{\text{p}}$ is deposited at time $t_{w(p-1)}$ and (b) second printing stage where $\text{L}_{\text{p}}$ is deposited on top of $\text{L}_{\text{p-1}}$ at time $t_{w(p)}$. Weld sites on either side of the layer-layer interface are denoted $\mathtt{t_{p-1}}$ and $\mathtt{b_{p}}$, and layer midpoints are denoted $\mathtt{m_p}$ and $\mathtt{m_{p-1}}$. (c) Spatio-temporal temperature profile of two printing stages and (d) temperature evolution at sites indicated in (a) and (b). Print temperature is $T_N=250^\text{o}$C and ambient temperature is $T_a=95^\text{o}C$. Welding between layers begins at time $t_{w(p)}$ and the layer-layer interface $z=0$ approaches the glass transition temperature $T_g = 140^\text{o}$C at time $t_g^W=1$ s.  }
\label{fig:tempstages}
\end{figure*}

In FFF the welding behaviour is essentially a thermally-driven diffusive process \cite{Sun:2008}, and interdiffusion is limited as the melt rapidly cools towards its glass transition temperature \cite{Seppala:2016}. In addition, large shear rates in the nozzle deform and align the polymer microstructure prior to welding; it is suggested that this alignment can affect the diffusive behaviour at the weld line causing de-bonding \cite{Croccolo:2013}. The deformation induced by the FFF extrusion and deposition process, which involves a $90^\text{o}$ turn, has recently been investigated using a molecularly-aware model for a non-crystalline polymer melt \cite{McIlroy:2016}. In that paper both stretch and orientation of the polymer are incorporated using the Rolie-Poly constitutive equation \cite{Likhtman:2003} and the entanglement density is allowed to vary with the flow \cite{Ianniruberto:2014}. Flow through the nozzle followed by deposition into an elliptically-shaped layer induces complex, non-axisymmetric polymer configurations, with the polymer microstructure varying dramatically from the top to the bottom of the printed layer. This deformation significantly disentangles the polymer melt via convective constraint release \cite{Ianniruberto:1996} in an inhomogeneous way. 

Due to this deformation imposed by the FFF extrusion process, interdiffusion does not necessarily occur from an equilibrium state. Non-equilibrium molecular dynamics (NEMD) calculations of a diffusion tensor for relatively short polymer melts under both planar Couette and planar elongational flow show a significant enhancement of diffusion parallel to the flow direction \cite{Hunt:2009}. Recently, models that incorporate an anisotropic shear-rate-dependent friction coefficient, or mobility tensor, have been proposed  to successfully reproduce the dynamics of polymers under shear \cite{Uneyama:2011, Ilg:2011}. Suitably accounting for flow-induced friction-reduction effects is also required to quantitatively model uniaxial extensional data \cite{IannirubertoMacro:2015}. The simple dynamical model of polymers under shear with an anisotropic  mobility tenor of Uneyama {\em et al.} \cite{Uneyama:2011} is consistent with NEMD simulations and experimental data; polymer segment alignment is suggested as the main cause for the anisotropy of the diffusion tensor. However, this model is not expected to apply o flows other than planar shear, and does not capture the anisotropic relaxation dynamics of aligned polymers after flow cessation.  

Furthermore, due to the nature of the deposition flow \cite{McIlroy:2016} polymers on either side of the interface reside in different deformation environments. Thus, a mutual diffusion mechanism should be considered, in which the diffusion coefficient depends on the local composition of chain  mobility \cite{Brochard:1983, Kramer:1984}. In particular, the theory of Kramer {\em et al.}  \cite{Kramer:1984}, suggesting that mutual diffusion is controlled by the mobility of faster moving chains, can successfully describe experimentally measured diffusion coefficients \cite{Composto:1986, Jordan:1988, Zhao:2007}.

In this paper, we investigate the post-extrusion diffusive behaviour at the weld between two printed layers of an amorphous (non-crystalline) polymer. We use the procedure developed by McIlroy \& Olmsted \cite{McIlroy:2016} to calculate the polymer conformation tensor and corresponding disentanglement that is induced by the extrusion process. We then introduce a spatio-temporal temperature profile to examine how this deformation relaxes in the weld region between two layers whilst cooling. In particular, we study how the structure of the weld evolves, how entanglements recover and calculate an interpenetration distance that incorporates both anisotropic and mutual diffusion effects. Finally, we address how these weld properties are affected by molecular weight, nozzle shear rate and print temperature.

\section{FFF Model}

\subsection{Ideal Extrusion Process}

The solid filament feedstock is melted within the nozzle. Recently Mackay {\em et al.} developed a model that solves an approximate energy balance to correlate the maximum feed velocity with the print temperature $T_N$ \cite{Mackay:2017}. In a frame moving with the nozzle, the melt exits the nozzle in steady state at mass-averaged speed $U_N$. It is then deposited onto the build surface, where the material must speed up and deform to make a $90^\text{o}$ turn. The build plate moves horizontally in relation to the nozzle at mass-averaged speed $U_L$. Assuming mass conservation and deformation into a cylindrical deposited filament, the two speeds are related by
\begin{equation}
 \pi R^2 U_N = \frac{\pi R H}{2} U_L,
 \label{eq:massconservation}
\end{equation}
where $R$ is the nozzle radius and $H$ is the thickness of the deposition. Typically $H<2R$, so that the deposited filament is elliptically shaped. The deposited filament then cools from the print temperature $T_N$ towards the glass transition temperature $T_g$. Typical model parameters are detailed in the Appendix (Table \ref{tab:speeds}).

For comparison with Ref. \cite{Seppala:2016}, we consider an extrusion process that deposits a single filament (or layer) in the $xy$-plane. Subsequent filaments are deposited on top of the previously-deposited filament to create a vertical printed `wall' in the $z$-direction (Fig. \ref{fig:process}). Due to this geometry we use the term `layer' to refer to a single deposited filament; note that some authors use `layer' to refer to the planar geometry of the build.  Figs. \ref{fig:tempstages}a and b illustrate these two key stages of FFF printing; layer $\mathtt{L_{p-1}}$ is deposited at time $t_{w(p-1)}$, followed by layer $\mathtt{L_{p}}$ on top of $\mathtt{L_{p-1}}$ at time $t_{w(p)}$. During the second stage the previously-printed layer heats up and a weld forms between the adjacent layers. 

The temperature profile $T(t, {\bf x})$ of the two layers drives the welding across the layer-layer interface $z=0$ in the region of the polymer size ($\pm R_g$); we denote weld sites either side of the interface to be $\mathtt{t_{p-1}}$ (at the top of $\mathtt{L_{p-1}}$) and $\mathtt{b_{p}}$ (at the bottom of $\mathtt{L_{p}}$), respectively. Inter-diffusion between the layers occurs until the weld temperature reaches $T_g$ at time $t_g^W$. 

\subsection{Temperature Profile}
\label{sec:temp}

In typical FFF processes, the nozzle is fixed at temperature $T_N$, the build plate is held at the ambient temperature $T_a$ (usually just below $T_g$), and printing occurs within an oven. The temperature profile of the oven is inhomogeneous due to the moving nozzle, which is continually accelerating and decelerating according to the print geometry and generating complex air-flow patterns. Thus, heat flow through the layers and exchange with the air is a very complicated problem. Typically, the boundary conditions at the layer-layer and layer-air interfaces are determined by a combination of convection, conduction and radiation. However, the heat transfer coefficient to describe this cooling process is not well understood. 

We neglect temperature variations in the $x$-direction and model the one-dimensional temperature profile $T(t,z)$ across two layers ($z\in [-H,H]$, where $z=0$ is the layer-layer interface and $z=H$ is the layer-air interface) via the one-dimensional heat equation
\begin{equation}
\frac{dT}{dt} = \alpha(T) \frac{\partial^2 T}{\partial z^2}.
\label{eq:temp}
\end{equation}
This gives the temperature evolution through the centre $(x=0)$ of the two layers (Fig. \ref{fig:tempstages}b). For polycarbonate, which is a typical printing material, the thermal diffusivity $\alpha$ has a linear temperature dependence \cite{Zhang:2002} of the form
\begin{equation}
 \alpha(T) = \alpha_0(1 - B T),
 \label{eq:alpha}
\end{equation}
for constants $\alpha_0, B$ and $T_g=140^\text{o}$C. Thermal diffusivity changes by $\sim 30$\% between the melt at $T_N$ and the solid at $T_a$ (see Appendix, Table \ref{tab:polycarbonate}). We solve Eqs. \ref{eq:temp}-\ref{eq:alpha} via an explicit finite-differencing scheme with the following boundary conditions. 

We assume Dirichlet boundary conditions at the interfaces such that 
\begin{equation}
 T(t,z=\pm H) = T_a,
\end{equation}
and the initial temperature profile is a step function, 
\begin{equation}
 T(t=0,z) = \begin{cases} T_N, \quad z > 0, \\
                          T_a, \quad  z < 0.
            \end{cases}
            \label{eq:initialtemp}
\end{equation}

\begin{figure}[t]
\centerline{\includegraphics[width=8cm]{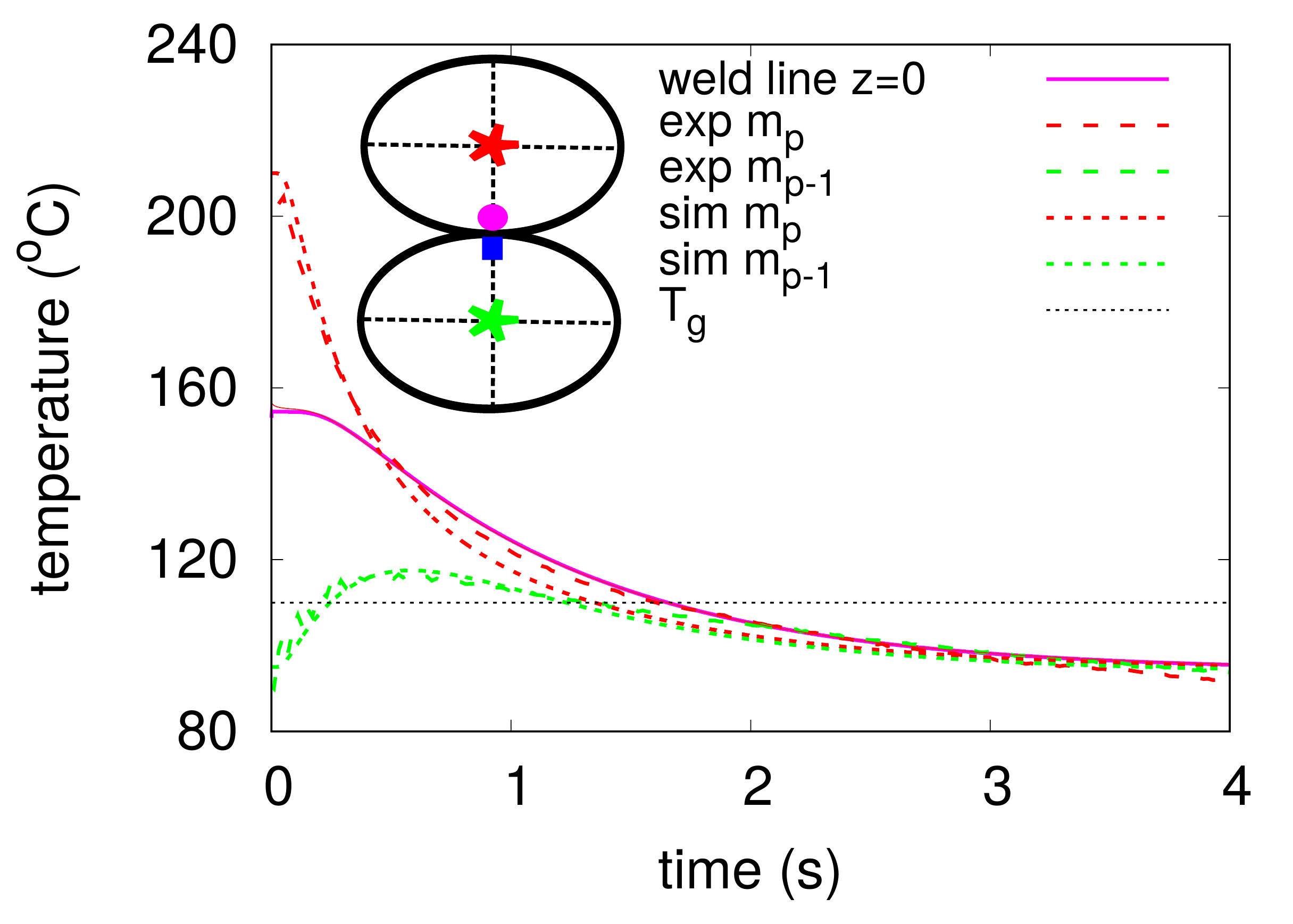}}
\caption{Temporal temperature evolution at layer midpoints $\mathtt{m_p}$ and $\mathtt{m_{p-1}}$ during the second stage of printing compared to experimental infrared measurements of the surface temperature of a printed ABS `wall' \cite{Seppala:2016}; ABS has similar thermal diffusivity to polycarbonate, a smaller $T_g=110^\text{o}$C, a smaller nozzle temperature $T_N=210^\text{o}$C and equivalent ambient temperature $T_a=95^\text{o}C$. The temperature evolution of the weld line is difficult to measure using infrared so only the prediction is plotted.}
\label{fig:ABScomp}
\end{figure}

Although Eqs. \ref{eq:temp}-\ref{eq:initialtemp} give a crude approximation for cooling, this thermal protocol mimics experimental infrared measurements of the surface temperature of a printed ABS `wall' \cite{Seppala:2016}, which has similar thermal diffusivity to polycarbonate. 
In particular, Fig. \ref{fig:ABScomp} shows that the one-dimensional model well describes the infrared-measured temperature evolution at layer midpoints $\mathtt{m_p}$ and $\mathtt{m_{p-1}}$ during the second stage of printing. The temperature at the weld line is hard to measure using infrared imaging due to the curvature of the surface; in Ref. \cite{Seppala:2016} it is determined by taking an average of the temperature evolution at $\mathtt{m_p}$ and $\mathtt{m_{p-1}}$. Arguably, the one-dimensional calculation shown in Fig \ref{fig:ABScomp} gives a more accurate description of the weld line temperature evolution, although complex cooling dynamics during deposition are neglected. The idealised wall geometry used here also does not capture the effects of adjacent layers in the $xy$-plane on the temperature profile. 


Figs. \ref{fig:tempstages}c shows the spatio-temporal temperature profile $T(t,z)$ predicted by Eqs. \ref{eq:temp}-\ref{eq:initialtemp} for polycarbonate properties (see Table \ref{tab:polycarbonate}) - the material we will consider henceforthly as in Ref. \cite{McIlroy:2016}. Fig. \ref{fig:tempstages}d shows the temporal temperature evolution of sites at the top $\mathtt{t}$, middle $\mathtt{m}$ and bottom $\mathtt{b}$ of each layer. We use a log-linear scale for comparison with later results. We choose $T_N=250^\text{o}$C and $T_a=95^\text{o}$C, as in typical FFF systems. Welding at the interface $z=0$ is governed by the temperature evolution and material dynamics on either side of the weld line at weld sites $\mathtt{t_{p-1}}$ and $\mathtt{b_{p}}$. These sites are chosen such that
\begin{equation}
\mathtt{t_{p-1}}- \mathtt{b_{p}}= 2R_g,
\label{eq:weldsites}
\end{equation}
where $R_g \sim 10$ nm is the radius of gyration of an individual polymer. In particular, Fig \ref{fig:tempstages}d shows how layer-air interface at $\mathtt{t_{p-1}}$ rapidly cools below $T_g$ during stage 1. At $t_{w(p)}$ the next layer is deposited, creating a layer-layer interface, and instantly heats up weld site  $\mathtt{t_{p-1}}$; the second weld site $\mathtt{b_{p}}$ instantly cools by the same degree. During stage 2, the layer-layer interface cools much slower than the layer-air interface reaching $T_g$ in approximately $t_g^W=1$ s. We will study how this weld line temperature evolution affects the relaxation of the printing material at weld sites $\mathtt{t_{p-1}}$ and $\mathtt{b_{p}}$, and consequently the characteristics of the weld.

\subsection{Polymer Dynamics}
\label{sec:RoliePoly}

We describe the polymer microstructure using a modified version Rolie-Poly model \cite{Likhtman:2003} that includes flow-induced changes in the entanglement fraction, as in Ref. \cite{McIlroy:2016}. Essentially the Rolie-Poly model is a variation of the standard Doi-Edwards tube model for linear entangled polymer networks, which approximates the more powerful but unwieldy microscopic GLaMM model \cite{Graham:2003} to provide a simple one-mode constitutive equation for the stress tensor (see Appendix \ref{sec:appendix_model} for more details). 

Since surrounding chains restrict transverse motion in a melt, a polymer chain is restricted to a tube-like region. This tube represents topological constraints due to entanglements \cite{DoiEdwards:1986}. At equilibrium, the entanglement number of a melt is related to the molecular weight $M_w$ via
\begin{equation}
 Z_{eq} = \frac{M_w}{M_e},
\end{equation}
where $M_e$ is the molecular weight between entanglements (see Table \ref{tab:polycarbonate}). Motion of a chain along the contour of the tube is unhindered by topological constraints and is known as reptation. 

The polymer microstructure can be parametrised by a conformation tensor 
\begin{equation}
 {\bf A} = \frac{ \langle {\bf RR} \rangle }{3R_g^2},
\end{equation}
for end-to-end vector ${\bf R}$ and radius of gyration $R_g$, which satisfies the Rolie-Poly equation \cite{Likhtman:2003}
\begin{equation}
\begin{split}
 \frac{D {\bf A}}{Dt} &= {\bf K} \cdot {\bf A} + {\bf A} \cdot {\bf K}^T - \frac{1}{\tau_d(T,\dot{\gamma})} ({\bf A}-{\bf I}) \\ &- \frac{2}{\tau_R(T)} \left( 1 - \sqrt{\frac{3}{\text{tr}{\bf A}}}\right) \left( {\bf A} + \beta \sqrt{\frac{\text{tr}{\bf A}}{3}} ({\bf A}-{\bf I}) \right),
\label{eq:Rolie-Poly}
\end{split}
\end{equation}
where $\frac{D}{Dt} = \frac{\partial}{\partial t} + ({\bf u}\cdot\nabla)$ is the material derivative for fluid velocity ${\bf u}$, ${\bf K} = \nabla_{\alpha \beta} u_\alpha$ is the velocity gradient tensor and $\text{tr}{\bf A}$ denotes the trace of tensor $\bf A$. The first term in Eq. \ref{eq:Rolie-Poly} described how chains become stretched and oriented in the flow field, whereas the last two terms define two relaxation mechanisms: reptation along the tube and Rouse relaxation of the tube stretch, respectively. 

The convective constraint release (CCR) mechanism, where the motion of neighbouring tubes can release a topological constraint, is controlled by the parameter $\beta$. In fast flow conditions, CCR together with alignment can lead to a flow-induced decrease in the entanglement fraction $\nu = Z/Z_{eq}$. Thus, we incorporate flow-induced disentanglement via the recent kinetic equation of Ianniruberto \cite{Ianniruberto:2014}:
\begin{equation}
 \frac{d\nu}{dt} = - \beta \left(  {\bf K}: {\bf A} - \frac{1}{\text{tr}{\bf A}} \frac{d}{dt} (\text{tr}{\bf A}) \right) \nu + \frac{1-\nu}{\tau_d^{eq}(T)}.
 \label{eq:nu}
\end{equation}
Entanglement loss can be modified by changing $\beta$ (see Ref. \cite{McIlroy:2016}) and entanglements are regained via curvilinear diffusion along the tube i.e. reptation. 

At equilibrium, the relaxation times depend on temperature via the typical Williams-Landel-Ferry (WLF) equation \cite{Williams:1955}: the reptation time $\tau_d^{eq}$ governs the orientation of the tube:
\begin{equation}
 \tau_d^{eq}(T) = \tau_d^0 \exp \left(\frac{-C_1(T-T_0)}{T+C_2-T_0} \right),
 \label{eq:taudeq}
\end{equation}
and the Rouse time $\tau_R^{eq}$ governs the relaxation of the tube stretch:
\begin{equation}
 \tau_R^{eq}(T) = \tau_R^0 \exp \left(\frac{-C_1(T-T_0)}{T+C_2-T_0} \right).
 \label{eq:tauReq}
\end{equation}
Here $C_1$ and $C_2$ are the WLF constants, $T_0$ is the reference temperature and $\tau_d^0$ and $\tau_R^0$ are the reptation and Rouse time at $T_0$ given by \cite{Likhtman:2002}
\begin{subequations}
\label{eq:taud0}
\begin{align}
 \tau_R^0 &= \tau_e^0 Z_{eq}^2, \\
 \tau_d^0 &= 3 \tau_e^0 Z_{eq}^3 \left(1 - \frac{3.38}{\sqrt{Z_{eq}}} + \frac{4.17}{Z_{eq}} - \frac{1.55}{\sqrt{Z_{eq}}^3} \right),
 \end{align}
\end{subequations}
respectively, where $\tau_e^0$ is the Rouse time of one entanglement segment at $T_0$ (see Table \ref{tab:polycarbonate} ). 

To incorporate the anisotropic nature of polymers in flow, the reptation time is modified according to \cite{Ianniruberto:2014}
\begin{equation}
\frac{1}{\tau_d(T,\dot{\gamma})} = \frac{1}{\tau_d^{eq}(T)} + \beta \left ({\bf K}:{\bf A} - \frac{1}{\text{tr}{\bf A}} \frac{d \text{tr}{\bf A}}{dt} \right).
\label{eq:taud}
\end{equation}
where the temperature-dependence of the equilibrium reptation time is given by Eq. (\ref{eq:taudeq}). Thus, polymers that are more aligned (and therefore partially disentangled) can relax faster at a rate proportional to CCR. The Rouse time in Eq. \ref{eq:Rolie-Poly} does not depend on the flow.

Often the entanglement number is written as
\begin{equation}
 Z_{eq} \approx \frac{3\tau_d^{eq}}{\tau_R^{eq}}.
\end{equation}
However, this relation is strictly only true for $Z_{eq} > 100$. Since printing materials typically have fewer entanglements, it is important to acknowledge that from Eq. \ref{eq:taud0}b the reptation time scales as
\begin{equation}
 \tau_d^0 = \frac{3}{11} \tau_e^0 Z_{eq}^{7/2}, 
 \label{eq:tdscaling}
\end{equation}
in the range $6<Z_{eq}<50$, rather than $Z_{eq}^3$. This is consistent with experiments that observe a 3.4 scaling \cite{DoiEdwards:1986}.

Flow through the nozzle is characterised by the equilibrium mass-averaged nozzle Weissenberg number by
\begin{equation}
 \overline{Wi} = \frac{U_N}{R} \tau_d^{eq}(T_N).
 \label{eq:WiN}
\end{equation}
For $\overline{Wi}>1$, the flow rate exceeds the characteristic relaxation time and can thus induce significant departure from the equilibrium polymer configuration. Similarly the equivalent average Rouse Weissenberg number is given by
\begin{equation}
 \overline{Wi}_R = \frac{U_N}{R} \tau_R^{eq}(T_N),
 \label{eq:WiNR}
\end{equation}
and signifies stretching of the tube. 


\section{Modelling the Weld Region}

\subsection{Initial Condition}

\begin{figure*}[t]
\centerline{\includegraphics[width=17cm]{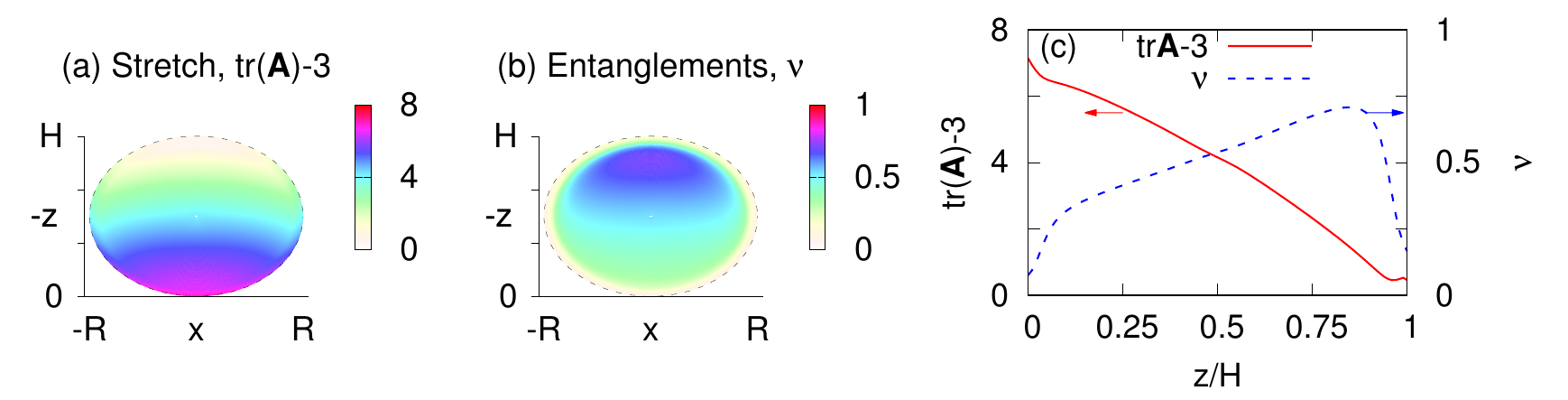}}
\caption{(a) Initial stretch profile $\text{tr}{\bf A}$ and (b) initial disentanglement fraction $\nu$ induced by the extrusion process across a printed layer. (b) Quantitative plot of the stretch and $\nu$ along the $z$-axis ($x=0$). The model parameters are $T_N=250^\text{o}$C, $\overline{Wi}=2$, $Z_{eq}=37$ and $\beta=0.3$. The polymer structure is highly stretched and disentangled at the bottom compared to the top of the layer. }
\label{fig:initialstretch}
\end{figure*}

Successful welding depends on the how the polymer melt interdiffuses and re-entangles across the layer-layer interface. To understand the interdiffusion process, we first must quantify the microstructure induced by the extrusion and deposition process. This configuration will provide an initial condition to calculate the evolution of the polymer structure at the weld. 

During deposition the polymer melt must deform to make the $90^\text{o}$ turn and transform into the elliptical geometry. Ref. \cite{McIlroy:2016} calculates the deformation $\bf A$ imposed by the extrusion and deposition process using the Rolie-Poly model (Sec. \ref{sec:RoliePoly}). The model parameters are $T_N=250^\text{o}$C, $\overline{Wi}=2$, $Z_{eq}=37$ and $\beta=0.3$. 

In particular, Fig. \ref{fig:initialstretch}a shows the stretch across the elliptical cross section of the deposit. There is a distinct gradient in the stretch from the top to the bottom of the layer, with the bottom half becoming much more stretched due to the stretching of the free surface during the $90^\text{o}$ turn. Due to polymer realignment during the deposition flow, the melt becomes significantly disentangled across the layer (Fig. \ref{fig:initialstretch}b).

\subsection{Dynamics at the Weld}

To determine how the polymer microstructure evolves at the weld after deposition, we solve the modified Rolie-Poly Eqs. \ref{eq:Rolie-Poly} and \ref{eq:nu} under zero flow conditions (i.e. ${\bf K}={\bf 0}$). The initial condition is calculated by the procedure in Ref. \cite{McIlroy:2016} (e.g. Fig. \ref{fig:initialstretch}). We then calculate the relaxation process at the weld sites $\mathtt{t_{p-1}}$ and $\mathtt{b_p}$ on either side of the weld line during the two stages of printing. During stage 1, $\mathtt{t_{p-1}}$ is at a layer-air interface, whereas during stage 2, $\mathtt{t_{p-1}}$ and $\mathtt{b_p}$ form a layer-layer interface (Fig. \ref{fig:relaxvisual}). The temperature dependence of the reptation and Rouse times (Eqs. \ref{eq:taudeq} and \ref{eq:tauReq}) is determined by the temperature profile at the weld calculated in Sec. \ref{sec:temp}. The model parameters are set to $T_N=250^\text{o}C$, $Z_{eq}=37$ and $\overline{Wi}=2$, which are typical for polycarbonate printing material. The CCR parameter is set to $\beta=0.3$ as in Ref. \cite{McIlroy:2016}.


\section{Evolution of Disentangled Weld Structure}

\begin{figure*}[p!]
\centerline{\includegraphics[width=18cm]{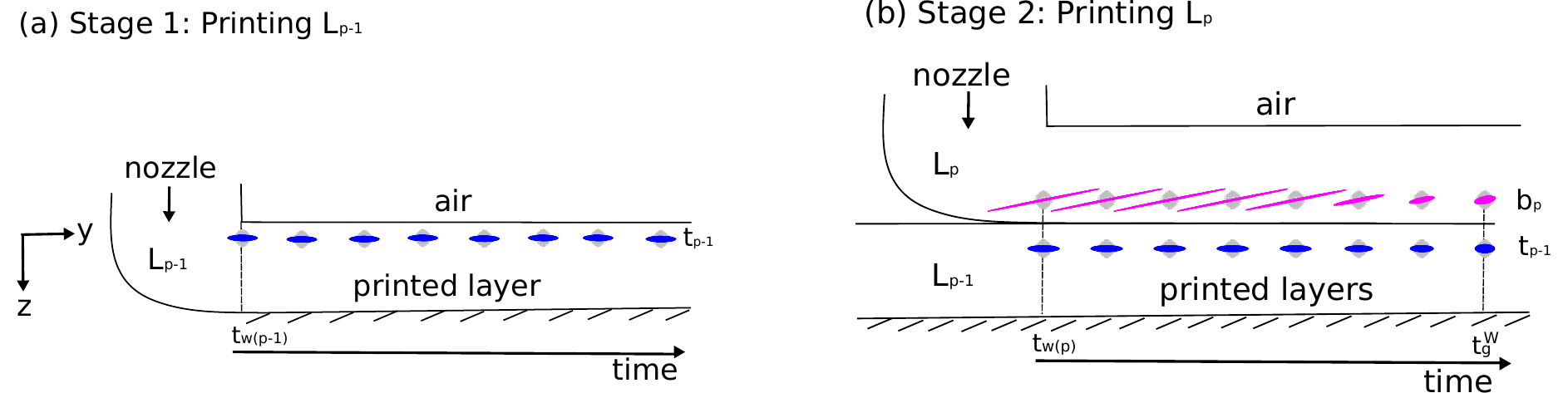}}
\caption{Elliptical representation of polymer conformation tensor ${\bf A}$ in the $yz$-plane during two stages of printing. (a) Stage 1: Layer $\mathtt{L_{p-1}}$ is deposited at time $t_{w(p-1)}$ and welding site $\mathtt{t_{p-1}}$ at the layer-air interface rapidly cools below $T_g$ and so that the deformation induced by extrusion only relaxes slightly. (b) Stage 2: Layer $\mathtt{L_p}$ is deposited at time $t_{w(p)}$ creating layer-layer interface between weld sites $\mathtt{t_{p-1}}$ and $\mathtt{b_p}$ and welding begins. Polymer relaxes until temperature of the weld reaches the glass transition at time $t_g^W$. Model  parameters are $T_N=250^\text{o}$C,  $\overline{Wi}=2$, $Z_{eq}=37$ and $\beta=0.3$. The structure of the weld is anisotropic at $t_g^W$.}
\label{fig:relaxvisual}
\vfill 
\centerline{\includegraphics[width=16cm]{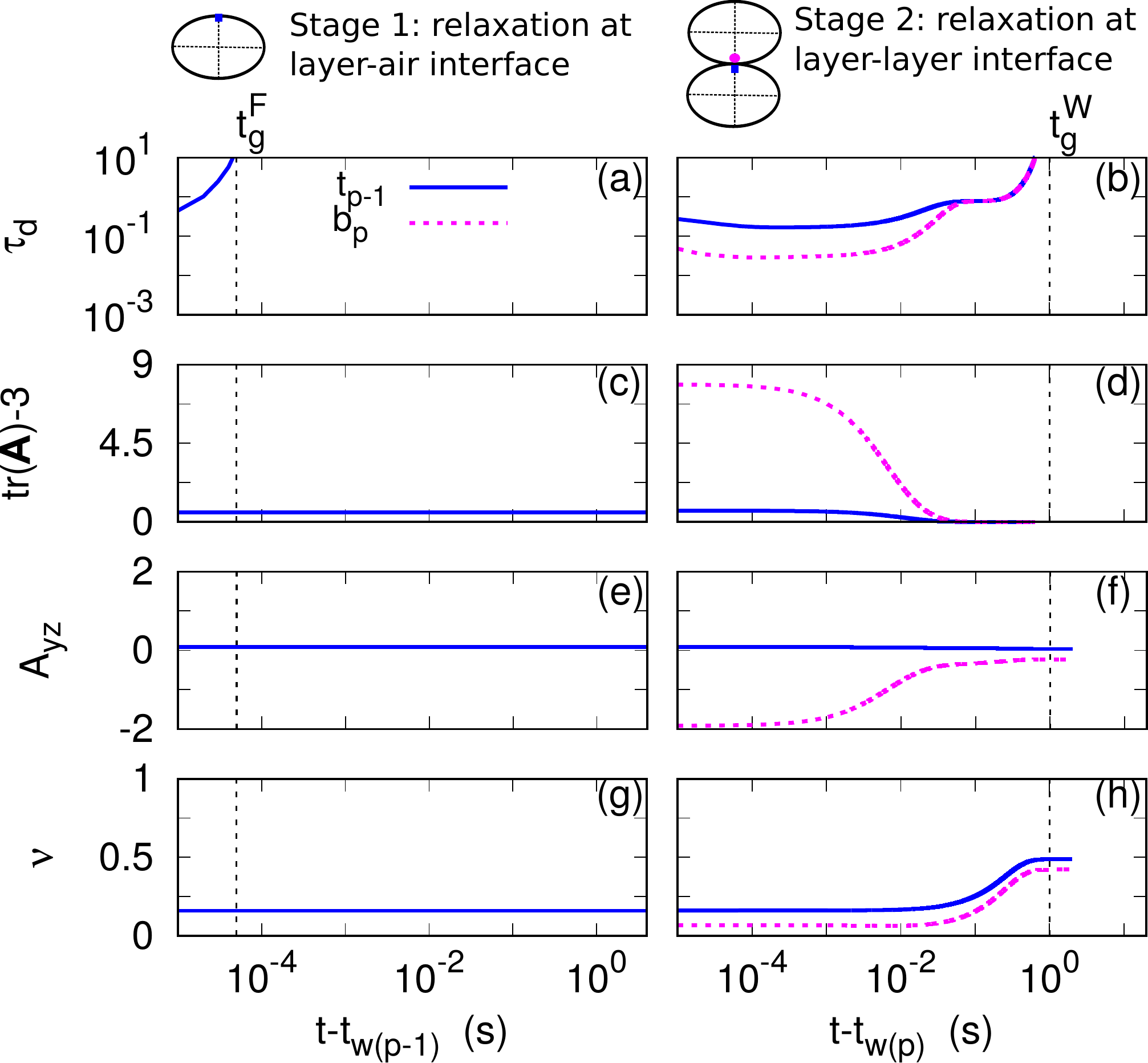}}
\caption{Time evolution of (a,b) the reptation time $\tau_d$ given by Eq. (\ref{eq:taud}) (c,d) the tube stretch $\text{tr}{\bf A}-3$, (e,f) the principle shear component $A_{yz}$, and (g,h) the entanglement fraction $\nu$  at weld sites $\mathtt{t_{p-1}}$ and $\mathtt{b_{p}}$ during two stages of printing. Dashed lines indicate the time at which the glass transition occurs at the free surface $t_g^F \approx 0.5 \mu$s and at the weld $t_g^W\approx 1$ s. Model parameters are $T_N=250^\text{o}$C,  $\overline{Wi}=2$, $Z_{eq}=37$ and $\beta=0.3$. The structure of the weld is anisotropic and disentangled at $t_g^W$ $(\nu \approx 0.5)$. }
\label{fig:relaxstages}
\end{figure*}

\begin{figure*}[t]
 \centerline{\includegraphics[width=16cm]{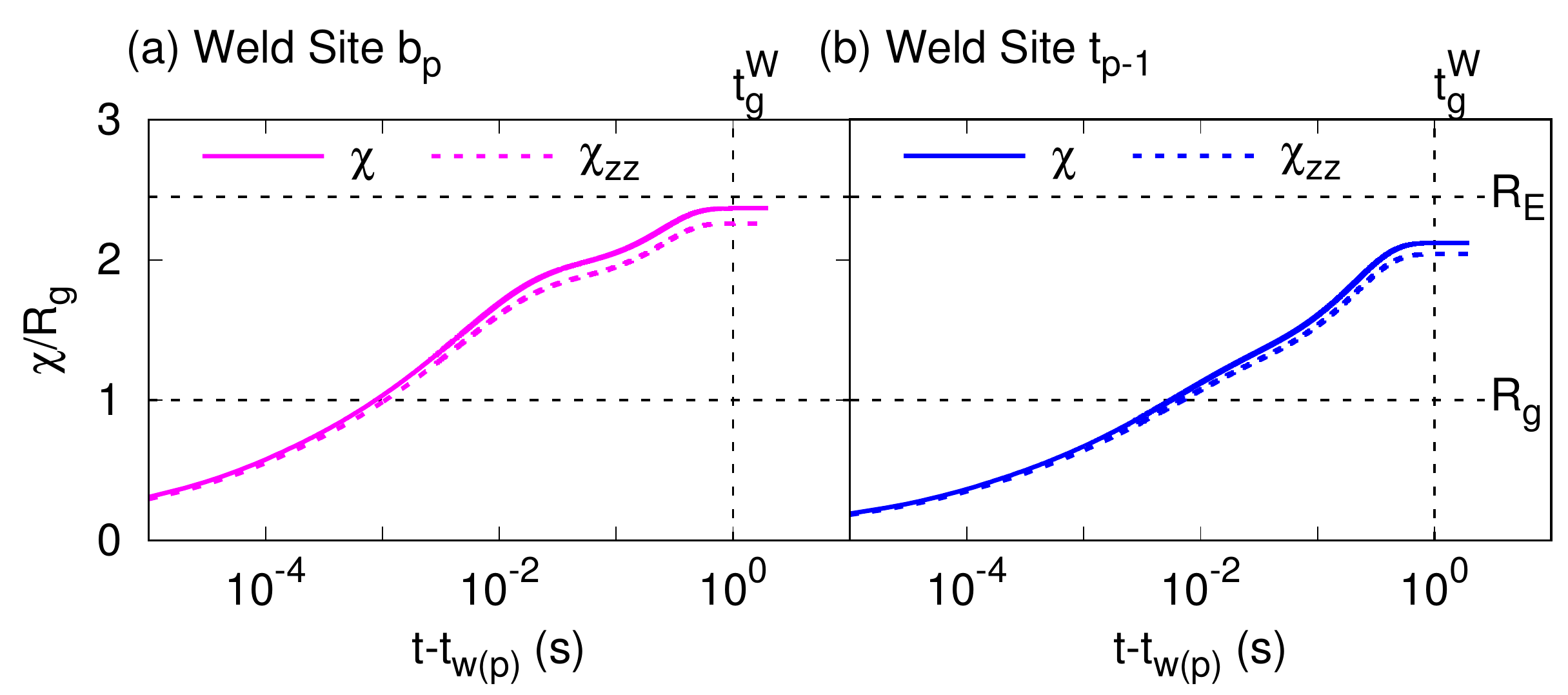}}
\caption{Inter-penetration distance calculated at weld sites (a) $\mathtt{b_p}$ and (b) $\mathtt{t_{p-1}}$. Isotropic welding given by $\chi$ (Eq. \ref{eq:weldthickness}) and anisotropic welding given by $\chi_{zz}$ (Eq. \ref{eq:anisowelding}). Dashed lines indicate $R_g$ and $R_E$ and $t_g^W \approx 1$ s. Model parameters are $T_N=250^\text{o}$C,  $\overline{Wi}=2$, $Z_{eq}=37$ and $\beta=0.3$. Inter-diffusion across the weld (in $z$-direction) is asymmetric and slowed by the anisotropic environment induced by extrusion.  }
\label{fig:anisowelding}
\end{figure*}


Fig. \ref{fig:relaxvisual} shows an elliptical representation of tensor ${\bf A}$ at weld sites $\mathtt{t_{p-1}}$ and $\mathtt{b_p}$; a sphere represents an undeformed polymer at equilibrium, whereas an ellipse represents a stretched and oriented polymer. The grey circles correspond to the equilibrium shape. We discuss the effects of changing the print speed and the CCR parameter $\beta$ on the relaxation dynamics in the Appendices \ref{sec:appendix_speed}, \ref{sec:appendix_CCR}. 

\underline{Stage 1 (Fig. \ref{fig:relaxvisual}a):} Layer $\mathtt{L_{p-1}}$ exits the nozzle at temperature $T_N$ and is deposited at time $t_{w(p-1)}$. The left-most ellipse represents the initial polymer configuration at weld site $\mathtt{t_{p-1}}$ induced by the deposition process. Since site $\mathtt{t_{p-1}}$ is exposed to the air it rapidly cools to the ambient temperature $T_a$ (Fig. \ref{fig:tempstages}d); the temperature of the free surface drops below $T_g$ in $t_g^F =0.5 \mu $s. Thus, the extrusion-induced deformation and corresponding disentanglement fraction do not have time fully relax and a non-equilibrium polymer configuration is locked into the weld site prior to the creation of the layer-layer interface.

\underline{Stage 2 (Fig. \ref{fig:relaxvisual}b):} Layer $\mathtt{L_p}$ exits the nozzle at temperature $T_N$ and is deposited on top of the cool layer $\mathtt{L_{p-1}}$ at time $t_{w(p)}$. This creates a layer-layer interface between weld sites $\mathtt{t_{p-1}}$ and $\mathtt{b_p}$. Again the left-most ellipses represent the initial polymer configuration at time $t_{w(p)}$. Each site has a different initial microstructure due to different degrees of deformation at the top and bottom of the layer during deposition, as well as the thermal history. In particular, the initial ellipse at $\mathtt{t_{p-1}}$ represents the deformation induced at the top of a layer that has been frozen in by the cooling of the layer-air interface during stage 1. On the other hand, the initial ellipse at $\mathtt{b_p}$ represents the deformation induced at the bottom of a layer at print temperature $T_N$, which is much larger due to the greater stretch around the outer corner. Since the layer-layer interface cools much slower than the layer-air interface (Fig \ref{fig:tempstages}d), the polymer has much longer to relax before $T_g$ is reached ($t_g^F \ll t_g^W$). Thus we see relaxation of the polymer at $\mathtt{t_{p-1}}$. Similarly, the larger deformation at $\mathtt{b_{p}}$ also relaxes with a similar temperature evolution, also reaching $T_g$ at time $t_g^W$. There is still insufficient time for the polymer to fully relax before the onset of the glass transition, so the weld region remain slightly anisotropic at $t_g^W$.

Fig. \ref{fig:relaxstages} shows how the reptation time $\tau_d$, the tube stretch $\text{tr}{\bf A}-3$, the principle shear component $A_{yz}$ and the entanglement fraction $\nu$ at the weld sites $\mathtt{t_{p-1}}$ and $\mathtt{b_{p}}$ evolve during the two stages of printing; note that $\mathtt{b_{p}}$ only exists once $\mathtt{L_p}$ has been deposited during the second stage of printing. We discuss these features in turn.

{\bf Reptation Time} (Fig. \ref{fig:relaxstages}a,b): After deposition of $\mathtt{L_{p-1}}$, the reptation time $\tau_d$ rapidly diverges (Fig \ref{fig:relaxstages}a), since the temperature at the layer-air interface $\mathtt{t_{p-1}}$ drops below $T_g$ in less than $100\mu$s. When the layer-layer interface is formed at time $t_{w(p)}$, heat transfer between the two layers causes the polymer at $\mathtt{t_{p-1}}$ to instantaneously become more mobile and the reptation time becomes finite (Fig \ref{fig:relaxstages}b). Despite having a similar temperature evolution, the two weld sites $\mathtt{t_{p-1}}$ and $\mathtt{b_p}$ have different reptation times due to the different degree of stretch induced by the extrusion process (Eq. \ref{eq:taud}).

{\bf Deformation Relaxation} (Fig. \ref{fig:relaxstages}c,d,e,f): During stage 1, Fig. \ref{fig:relaxstages}c,e shows that the initial tube stretch $\text{tr}{\bf A}$ and shear deformation $A_{yz}$ at the free surface $\mathtt{t_{p-1}}$ do not relax due to the diverging reptation time and persist to the second printing stage. Once $\mathtt{L_{p}}$ is deposited the increased temperature of the weld site allows relaxation of $\text{tr}{\bf A}$ and $A_{yz}$ (Fig. \ref{fig:relaxstages}d,f). 

Since $\overline{Wi}_R>1$, linear relaxation of the deformation at both $\mathtt{t_{p-1}}$ and $\mathtt{b_p}$ does not apply. Instead, the relaxation process is two-stage; the first relaxation mode is Rouse-like and governed by $\tau_R$. Once the tube length returns to the equilibrium value $\text{tr}{\bf A} = 3$, the usual reptation behaviour prevails and the reptation time $\tau_d$ becomes equivalent for both weld sites (Fig. \ref{fig:relaxstages}b), now depending only on the temperature evolution. Convective constraint release (parametrised by $\beta$) only contributes to relaxation whilst $\text{tr}{\bf A}>3$ and therefore only has a small effect on the relaxation dynamics during the first Rouse relaxation mode (see Appendix \ref{sec:appendix_CCR}).

Fig \ref{fig:relaxstages}d shows that the stretch of the tube at both $\mathtt{t_{p-1}}$ and $\mathtt{b_p}$ has sufficient time to relax prior to the onset of the glass transition at time $t_g^W$. In contrast, the relaxation of principle shear component $A_{yz}$ (Fig \ref{fig:relaxstages}f) is arrested by the glass transition. This anisotropy is particularly prominent at site $\mathtt{b_p}$. Thus, a non-equilibrium polymer orientation becomes `locked' into the weld region at $t_g^W$ so that the structure of the weld is slightly anisotropic.

{\bf Recovery of Entanglements} (Fig. \ref{fig:relaxstages}g,h): Finally, Fig. \ref{fig:relaxstages}h shows how the entanglement fraction $\nu$ at both sites recovers towards unity  during stage 2 according to Eq. (\ref{eq:nu}). Initially, the evolution of entanglements is governed by the tube stretch; whilst $\text{tr}{\bf A}>3$, $\nu$ remains constant since entanglements cannot be gained when the tube is retracting. Once the stretch has returned to equilibrium, entanglements recover at a rate determined by the reptation time $\tau_d^{eq}$, which depends only on the temperature evolution. Due to the arresting glass transition, entanglements are not able to fully recover and $\nu \approx 0.5$ at $t_g^W$ for both sites $\mathtt{t_{p-1}}$ and $\mathtt{b_p}$. Thus, the weld region is approximately 50\% less entangled than the equilibrium material, presumably yielding a lower mechanical strength compared to the equilibrium material. Since $\beta$ only affects entanglement recovery during the Rouse relaxation mode, we see see similar $\nu$ at $t_g^W$ for all $\beta$ (see Appendix \ref{sec:appendix_CCR}).

\section{Weld Inter-Penetration Thickness}

\begin{figure*}[t]
\includegraphics[width=19cm]{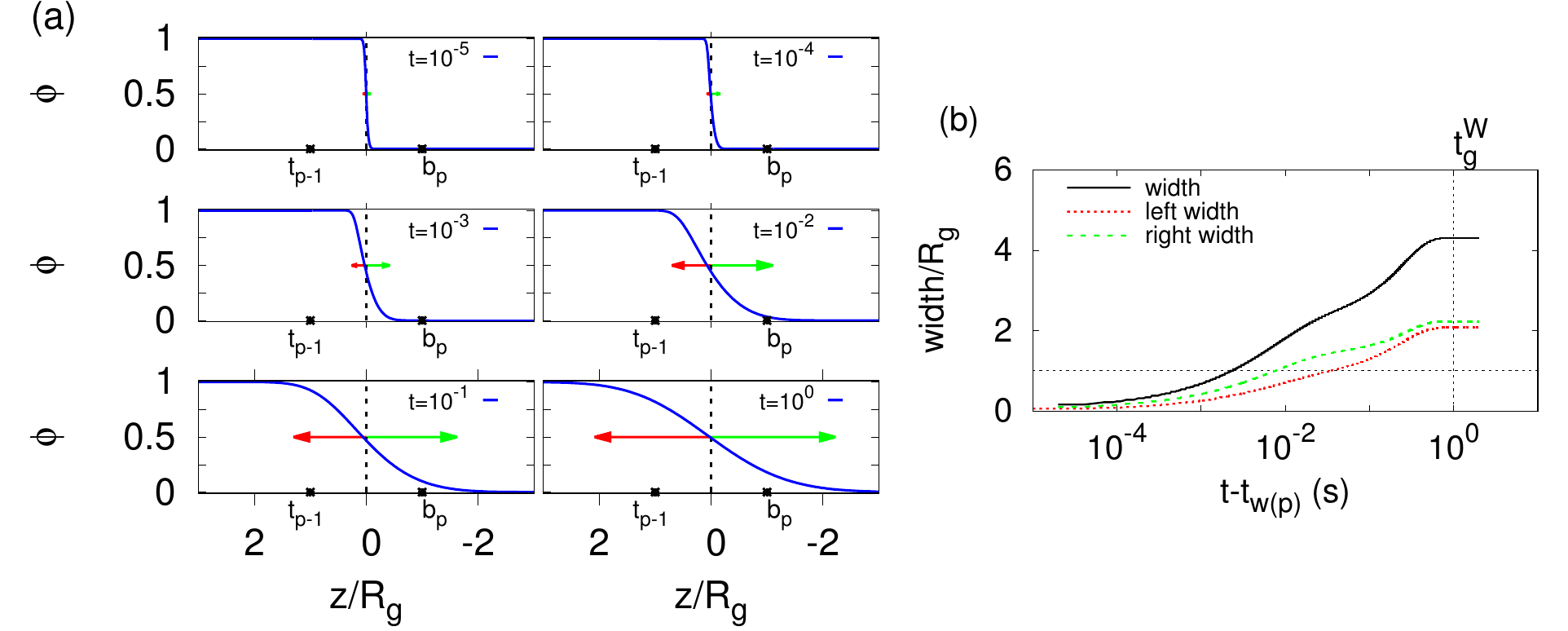}
\caption{(a) Evolution of the volume fraction $\phi$ profile (Eq. \ref{eq:phi}) in the region of the layer-layer interface $(z=0)$ and weld sites $\mathtt{t_{p-1}}$ and $\mathtt{b_p}$. Arrows indicate the interfacial width on the left {\em (red)} and on the right {\em (green)} of the interface. (b) Evolution of the interfacial width, and widths calculated to the left and right of the interface. Dashed line indicates $R_g$ and $t_g^W \approx 1$s. Model parameters are $T_N=250^\text{o}$C,  $\overline{Wi}=2$, $Z_{eq}=37$ and $\beta=0.3$. The weld thickness is asymmetric.  }
 \label{fig:phiprofile}
\end{figure*}

We now consider diffusion dynamics at the two weld sites $\mathtt{t_{p-1}}$ and $\mathtt{b_p}$ located at distances $\pm R_g$ from the weld line (Eq. \ref{eq:weldsites}). First we consider isotropic diffusion that is sped up by the relaxation of the tube stretch according to Eq. \ref{eq:taud}. We then include the effects that an anisotropic environment has on the diffusion direction. Finally, we incorporate inhomogeneous diffusivity across the chain due to the polymer diffusing into a different deformation environment as it crosses the weld line.

\subsection{Isotropic Welding Approximation}


The curvilinear diffusion coefficient along the contour length of the tube is given by \cite{DoiEdwards:1986}
\begin{equation}
D_c = \frac{k_B T}{N \zeta}
\end{equation}
where $N$ is the number of Kuhn steps along the path, $\zeta$ is the friction coefficient and $k_B$ is the Boltzman constant. The Kuhn length $b$ is the statistical length of a polymer segment and often represents chain stiffness \cite{DoiEdwards:1986}.

The curvilinear distance $\ell$ travelled by a polymer chain in time $t$ is then given by the Einstein relation for a one-dimensional random walk \cite{Wool:1981}
\begin{equation}
 \langle \ell^2 \rangle= 2 D_c t = \frac{L_c^2t}{\tau_d}.
\end{equation}
The tube contour length is defined by \cite{DoiEdwards:1986}
\begin{equation}
 L_c = \frac{Nb^2}{a_T},
\end{equation}
for Kuhn length $b$ and tube diameter 
\begin{equation}
a_T = b\sqrt{N_e},
\label{eq:tubediam}
\end{equation}
where $N_e$ is the number of Kuhn steps between entanglements; at equilibrium, $N_e= N/Z_{eq}$.

The 1D tube contour executes a random walk in 3D-space, leading to an interpenetration distance $\chi$ given by \cite{Wool:1981, deGennes:1981}
\begin{equation}
\begin{split}
\chi^2  &= \langle \ell^2 \rangle^{1/2} a_T, \\
       &= N b^2 \left( \frac{t}{\tau_d} \right)^{1/2}
\end{split}
\end{equation}
Thus, a polymer chain diffuses via a double random walk process. 

In the FFF problem the decreasing temperature progressively slows the motion. Thus, the interpenetration distance is calculated via the integral
\begin{equation}
 \frac{\chi}{R_g} = \left( 36 \int_{t_w}^{t_g^W} \frac{1}{\tau_d(T(t),\dot{\gamma}(t))} dt' \right)^{1/4},
 \label{eq:weldthickness}
\end{equation}
where $R_g = Nb^2/\sqrt{6}$ and the reptation time depends on both temperature and shear rate (Eq. \ref{eq:taud}). 

For polycarbonate printed at $T_N=250^\text{o}$C, Fig. \ref{fig:anisowelding} shows the interpenetration depth for polymers located at the weld: polymers located at $\mathtt{b_{p}}$ travel slightly further than polymers located at $\mathtt{t_{p-1}}$ due to a smaller reptation time, which is a result of the increased tube stretch at this weld site. In particular, for our model parameters $ \chi \approx  2R_g$ at $\mathtt{b_p}$ before the glass transition arrests diffusion at $t_g^W$. Usually, experiments find that bulk strength is achieved once diffusion of the order of $R_g$ has occurred \cite{Wool:1995} and molecular simulations suggest that only a few entanglement lengths are required \cite{Robbins:2013}. 

Despite a diffusion distance greater than $R_g$, we find that orientations do not fully relax during this time, thus creating an anisotropic weld structure (Fig. \ref{fig:relaxvisual}). The polymer must diffuse its end-to-end distance $R_E = \sqrt{6} R_g$ to fully escape its tube and relax to equilibrium.

\subsection{Anisotropic Welding Approximation}

\begin{figure*}[t]
 \centerline{\includegraphics[width=16cm]{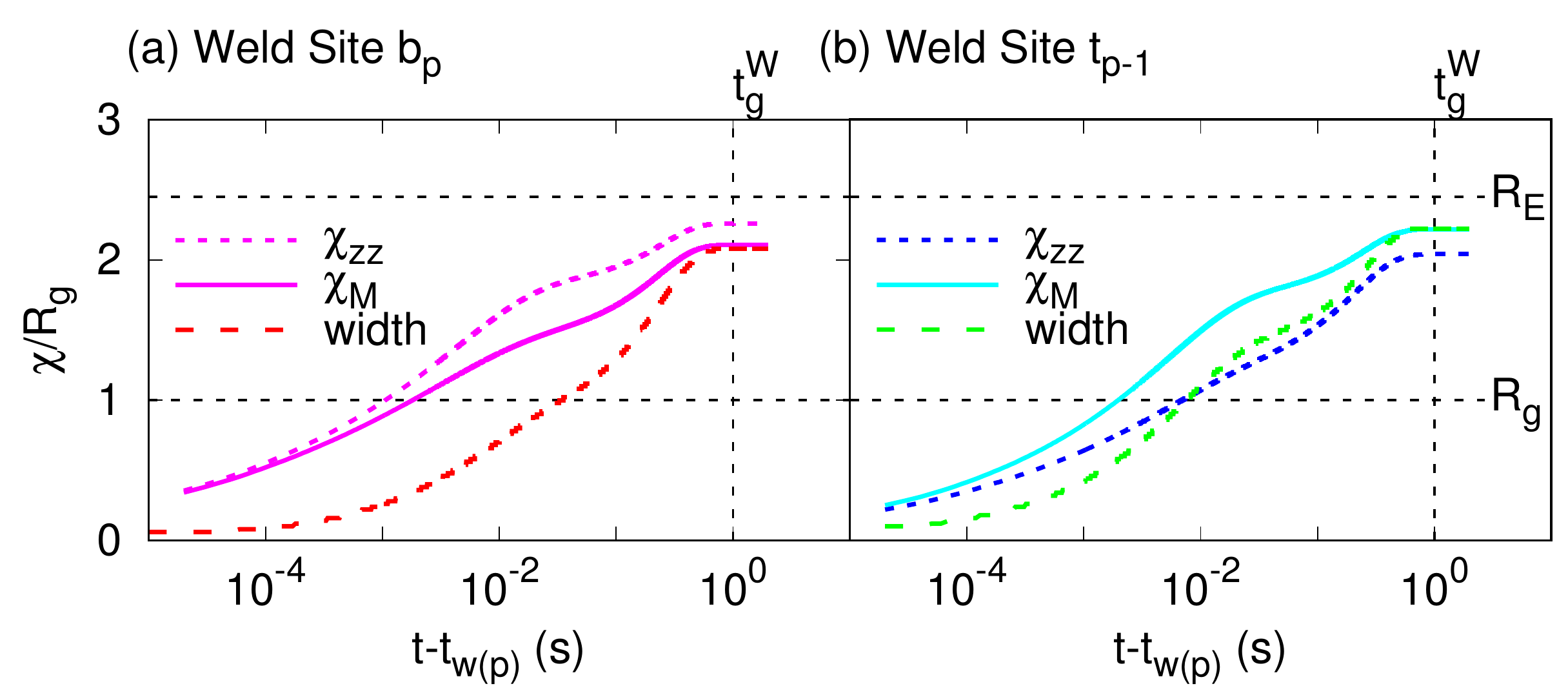}}
\caption{Time evolution of the interpenetration distance $\chi_{zz}$ incorporating anisotropic diffusion (Eq. \ref{eq:anisowelding}), and the interpenetration distance $\chi_M$ given by the mutual diffusion theory (Eq. \ref{eq:mutualwelding}) for polymers initially located at weld sites (a) $\mathtt{b_{p}}$ and (b) $\mathtt{t_{p-1}}$. Dashed lines indicate $R_g$ and $R_E$ and $t_g^W \approx 1$ s. Distances are compared to the the interfacial width determined by $\phi$ shown in Fig. \ref{fig:phiprofile}b. Model parameters are$T_N=250^\text{o}$C,  $\overline{Wi}=2$, $Z_{eq}=37$ and $\beta=0.3$. The weld thickness is asymmetric, with material diffusing further into more mobile environment (i.e. $\mathtt{t_{p-1}}$ towards $\mathtt{b_p}$). }
 \label{fig:mutualwelding}
\end{figure*}

Eq. \ref{eq:weldthickness} does not account for a preferred diffusion direction due to the local anisotropic structure of the polymer melt. 
Similar to the work of Ilg \& Kroger \cite{Ilg:2011}, we propose a time-dependent anisotropic diffusion tensor of the form
\begin{equation}
 {\bf D} = D_0 ({\bf I} + \eta ({\bf A}-{\bf I})),
\end{equation}
where $D_0 = Nb^2/\tau_d(t)$. The anisotropy parameter was found to be $\eta \simeq 1/3$ by molecular simulations of long chains \cite{Ilg:2011}. The interpenetration distance across the interface is then given by
\begin{equation}
 \frac{\chi_{zz}}{R_g} = \left( 36 \int_{t_w}^{t_g^W} \frac{1}{\tau_d(t)} \left( 1 + \eta (A_{zz}(t)-1) \right) dt' \right)^{1/4}.
 \label{eq:anisowelding}
\end{equation}

For polycarbonate printed at $T_N=250^\text{o}$C, Fig. \ref{fig:anisowelding} shows how the interpenetration depth across the weld line is reduced for anisotropic diffusion of polymers located at weld sites $\mathtt{t_{p-1}}$ and $\mathtt{b_p}$. Yet $\chi_{zz}$ is also larger than $R_g$, which according to molecular dynamics simulations \cite{Robbins:2013} suggests that bulk strength should be achieved in the weld region.

\subsection{Mutual Diffusion Approximation}

When a polymer molecule diffuses across the interface, it is not only affected by the anisotropy of its own structure, but also the anisotropy of the environment into which it diffuses. We consider polymers located either side of the interface of type $\mathtt{t}$ and $\mathtt{b}$ to signify the deformation at weld sites $\mathtt{{t}_{p-1}}$ and $\mathtt{b_p}$, respectively. Similar to the  work of Kramer {\em et al.} \cite{Kramer:1984},  we propose a mutual diffusion coefficient of the form
\begin{equation}
 {\bf D}^M (t,z) =  ( 1 - \phi(t,z)) {\bf D}^{\mathtt{b}}(t) + \phi(t,z) {\bf D}^{\mathtt{t}}(t),
 \label{eq:mutualdiff0}
\end{equation}
where
\begin{subequations}
\label{eq:tbdiffusion}
 \begin{align}
  {\bf D}^{\mathtt{b}}(t) & = \frac{ Nb^2}{\tau_d^\mathtt{b}(t)} \left ( {\bf I} + \eta({\bf A}^{\mathtt{b}}(t)-{\bf I}) \right), \\
 {\bf D}^{\mathtt{t}}(t) & = \frac{ Nb^2}{\tau_d^\mathtt{t}(t)} \left ( {\bf I} + \eta({\bf A}^{\mathtt{t}}(t)-{\bf I}) \right).
 \end{align}
\end{subequations}
The volume fraction occupied by type $\mathtt{t}$ polymers is denoted by $\phi$. Initially $\phi = 1$ at weld site $\mathtt{{t}_{p-1}}$ and $\phi=0$ at weld site $\mathtt{b_p}$. The volume fraction evolves according to
\begin{equation}
 \frac{\partial \phi}{\partial t} = \frac{\partial}{\partial z} \left(  D_{zz}^M(t,z) \frac{\partial \phi}{\partial z} \right),
 \label{eq:phi}
\end{equation}
where $D_{zz}^M$ is the $zz$-component of the mutual diffusion tensor governed by Eq. \ref{eq:mutualdiff0}. In this way a diffusing chain carries its mobility across the weld line and the diffusion coefficient depends on the local composition of mobility. 

For polycarbonate printed at $T_N=250^\text{o}$C, Fig. \ref{fig:phiprofile}a shows the evolution of $\phi$ during the relaxation process. The weld is formed due to polymers diffusing across the interface. Assuming that a weld is formed in the region $2 \% < \phi < 98\%$, Fig. \ref{fig:phiprofile}b. shows the evolution of the interfacial width of the weld region. We highlight the asymmetric nature of diffusion across the interface by plotting the width calculated to the left and the right of the interface. Asymmetry arises due to the different degree of deformation at weld sites $\mathtt{t_{p-1}}$ and $\mathtt{b_p}$. 

The Kramer model \cite{Kramer:1984} describes the mutual diffusion of two different molecular weight polymers \cite{Zhao:2007}. 
However, since the FFF problem discussed here involves a single molecular weight inter-diffusing between different-mobility environments, a polymer chain will inherit the relaxation characteristics of the environment into which it is diffusing. Thus, a similar approach can be taken to calculate the mutual interpenetration distance of a molecule located at weld site $\mathtt{t_{p-1}}$, namely $\chi_M$. That is
\begin{equation}
 \frac{\chi_M}{R_g} = \left( 36 \int_{t_w}^{t_g^W} \widetilde{ D}^M_{zz}(t)  dt' \right)^{1/4},
 \label{eq:mutualwelding}
\end{equation}
where the diffusion coefficient of this molecule is given by
\begin{equation}
 \widetilde{\bf D}^M(t) = \begin{cases}
                          \displaystyle{ \left( 1- \frac{\chi_M}{2R_g} \right) {\bf D}^{\mathtt{t}} + \frac{\chi_M}{2R_g} {\bf D}^\mathtt{b}}&, \quad \chi_M < 2R_g, \\[10truept]
                          \displaystyle{{\bf D}^\mathtt{b}} &, \quad  \chi_M \ge 2R_g,
                          \end{cases}
                          \label{eq:mutualdiff}
\end{equation}
and $\chi_M$ parametrises how far the polymer has penetrated. ${\bf D}^\mathtt{t}$ and ${\bf D}^\mathtt{b}$ are given by Eq. \ref{eq:tbdiffusion}. A similar relation is used for molecules located at weld site $\mathtt{b_p}$. In this way the mutual diffusion coefficient of a chain is given by a weighted average of the two environments on either side of the interface that depends on the polymer's location.

Compared to $\chi_{zz}$ (Eq. \ref{eq:anisowelding}), Fig. \ref{fig:mutualwelding}a shows how diffusion of molecules initially located at weld site $\mathtt{b_p}$ is slowed down by diffusing into a slower moving environment for polycarbonate printed at $T_N=250^\text{o}$C. In contrast, the diffusion of molecules initially located at $\mathtt{t_{p-1}}$ is increased by diffusing into a faster-moving environment (Fig. \ref{fig:mutualwelding}b). For this mutual diffusion model, type $\mathtt{t}$ polymers ultimately diffuse further across the interface due to the increased mobility of diffusing into a faster environment. Thus, despite type $\mathtt{b}$ polymers initially being more mobile due to the degree of deformation, the type $\mathtt{t} $ polymers ultimately create a thicker interfacial width. This approach gives a final asymmetric interfacial width similar to that seen in Fig. \ref{fig:phiprofile}b for the Kramer model based on local composition (Eq. \ref{eq:phi}). 

Including the effects of mutual diffusion also yields $\chi_M > R_g$ at this print temperature. Thus, for this case the mechanical strength of the weld region is limited the molecular structure (and entanglement fraction) of the weld region itself (Fig \ref{fig:relaxstages}), rather than the interpenetration depth of the polymer. 


\begin{figure*}[t]
 \centerline{\includegraphics[width=18cm]{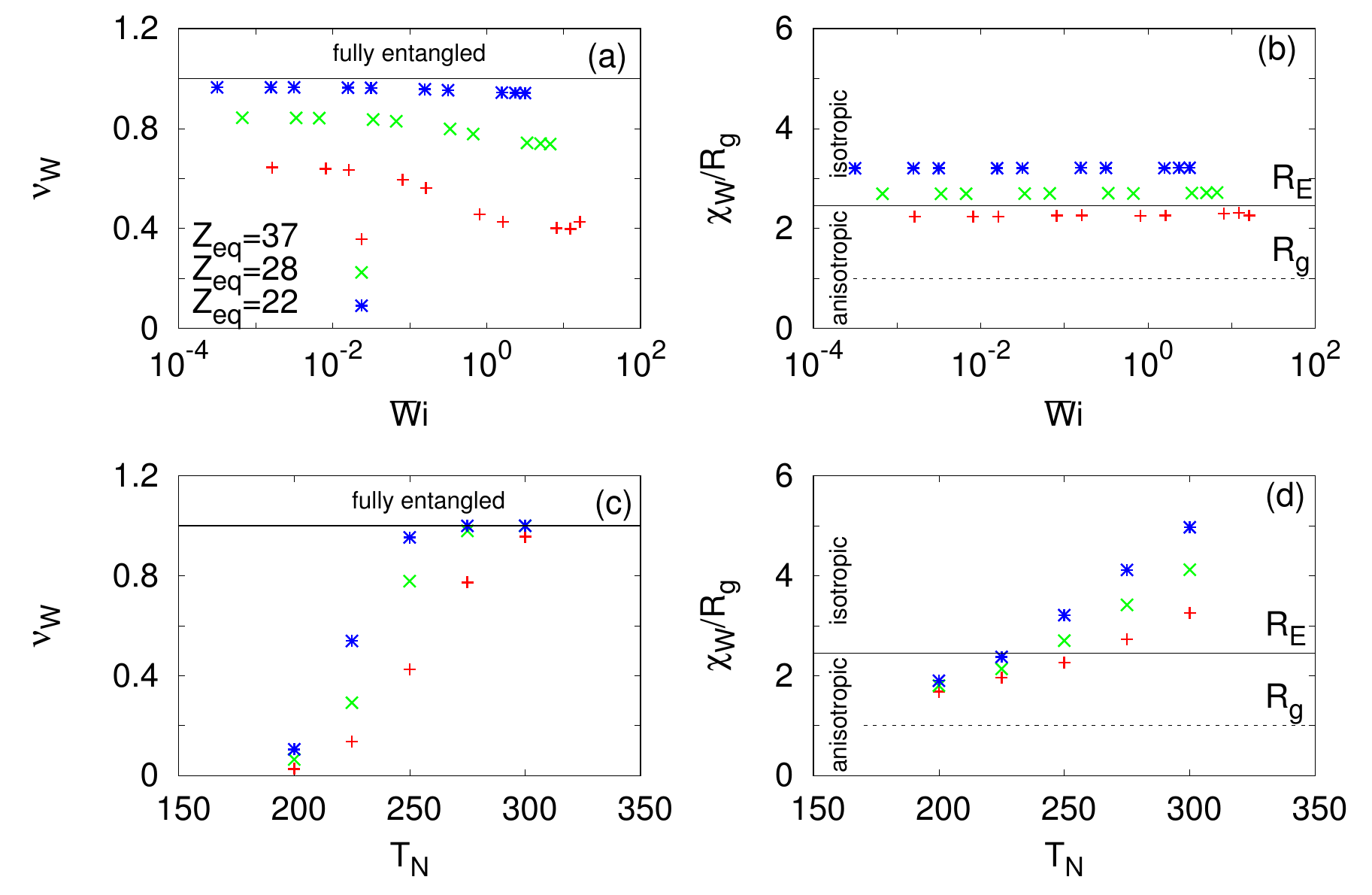}}
\caption{Weld characteristics (a,c) final weld entanglement fraction $\nu_W$ and (b,d) final weld thickness $\chi_W$ measured at $t_g^W$ for (a,b) a range of Weissenberg numbers $\overline{Wi}$ at $T_N=250^\text{o}$C and (c,d) a range of print temperatures $T_N$ at the slow print speed $U_N=10$ mm/s. For $\chi_W<R_E = \sqrt{6} R_g$ the weld structure will be anisotropic. Equilibrium entanglement numbers are $Z_{eq}=37,28$ and 22, similar to typical printing materials.}
\label{fig:weld}
\end{figure*}

\section{Controlling Weld Strength}

\subsection{Final Weld Properties}

The mechanical properties of a material can be characterised by the plateau modulus $G_e$ \cite{DoiEdwards:1986} and the fracture toughness $G_c$ \cite{Sha:1999, Rottler:2002}. Both properties are proportional to the  molecular weight between entanglements, $M_e$:
\begin{subequations}
\label{eq:GeGc}
 \begin{align}
  G_e &\sim \frac{1}{M_e}, \\
  G_c &\sim \left(1 - \frac{M_e}{qM_w}\right)^2, 
\end{align}
\end{subequations}
for $q\approx 0.6$ \cite{Sha:1999}. Smaller $M_e$ results in a greater entanglement density and therefore increases both the bulk modulus and fracture toughness of a material.

For polycarbonate, the entanglement molecular weight in determined by 
\begin{equation}
 M_e = 1.156 a_T^2 \text{ g/mol},
 \label{eq:Me}
\end{equation}
where $a_T = 37.9$ \AA \space is the tube diameter (Eq. \ref{eq:tubediam}) and the pre-factor accounts for bond angle, characteristic ratio and monomer weight \cite{Mark:1996}.

At a welded interface the mechanical strength is determined by how many segments of length $a_T$ cross the interface \cite{deGennes:1989}. Thus, the mechanical strength of a weld is attributed to the weld thickness $\chi_W$, as well as the integrity of the entanglement network at the weld $\nu_W$. Bulk strength is expected for $\nu_W=1$ and $\chi_W/R_g>1$. For certain printing conditions we have seen that, although the weld thickness exceeds $R_g$, the ultimate structure of the weld is anisotropic with a weaker entanglement structure. Thus, reliable prediction of the weld properties from material properties and printing parameters is key to ensuring weld strength and advancing FFF technology.

The FFF extrusion model in \cite{McIlroy:2016} assumes an idealised deposition process during which there is no cooling or relaxation of the melt. In addition, the assumed temperature profile in the nozzle neglects inhomogeneities due to thermal diffusion and shear heating effects. Both of these factors may significantly affect the temperature and polymer structure at the weld site, and consequently affect the weld properties. Thus, it is important to validate $\chi_W$ against experimentally measured weld thicknesses, which can be challenging to measure reproducibly and reliably from FFF-printed parts. 

In this section we draw qualitative conclusions of the weld properties based on our model. First we calculate the final weld entanglement number $\nu_W$ and the interpenetration depth $\chi_W$ at time $t_g^W$ and examine how they vary with the equilibrium mass-averaged nozzle Weissenberg number (Eq. \ref{eq:WiN}), print temperature $T_N$ and molecular weight $Z_{eq}$. Finally, we suggest how these weld properties can increase the entanglement molecular weight in the weld region and therefore decrease the mechanical strength of a printed part. 
 
\subsection{Effect of Print Speed}
 
For a fixed print temperature $T_N$ and a range of typical entanglement numbers $Z_{eq}$, Figs. \ref{fig:weld}a and b show weld properties $\nu_W$ and $\chi_W$ for increasing $\overline{Wi}$. 
We observe a slight decrease in $\nu_W$ with $\overline{Wi}$ for the largest molecular weight $Z_{eq}=37$, whereas $\chi_W$ is independent of the nozzle shear rate for all three $Z_{eq}$ shown (Fig. \ref{fig:weld}b). Thus, the increased initial stretch imposed by larger shear rates in the nozzle is not large enough to influence the weld thickness by reducing the reptation time.  Melts with fewer entanglements are more mobile and are therefore able to diffuse further during the welding process, creating a thicker, more entangled weld structure. For $ \chi_W < R_E$ the weld structure will be anisotropic at the glass transition. Once the interpenetration depth surpasses $R_E$ we expect an isotropic weld structure.

\subsection{Effect of Print Temperature}

In contrast to Weissenberg number, we find that print temperature significantly affects welding behaviour (Figs. \ref{fig:weld}c,d). Both $\nu_W$ and $\chi_W$ increase as a function of $T_N$ since higher temperatures significantly speed up diffusion. Notably, even at low print temperature $T_N =200^\text{o}$C the weld thickness continues to surpass $R_g$, suggesting healing of the interface. However, the weld site remains almost fully disentangled in this case (Fig. \ref{fig:weld}c), so that bulk strength is unattainable. This is a consequence of the delay in entanglement recovery due to the relaxation of the tube stretch (Fig. \ref{fig:relaxstages}h). Moreover, since smaller molecular weights have shorter Rouse times, entanglement recovery begins earlier, which allows the formation of a more entangled weld. Only for $T_N \ge 300^\text{o}$C does the weld becomes fully entangled for all $Z_{eq}$.

\subsection{Effect of Molecular Weight}

\begin{figure*}[t!]
 \centerline{\includegraphics[width=18cm]{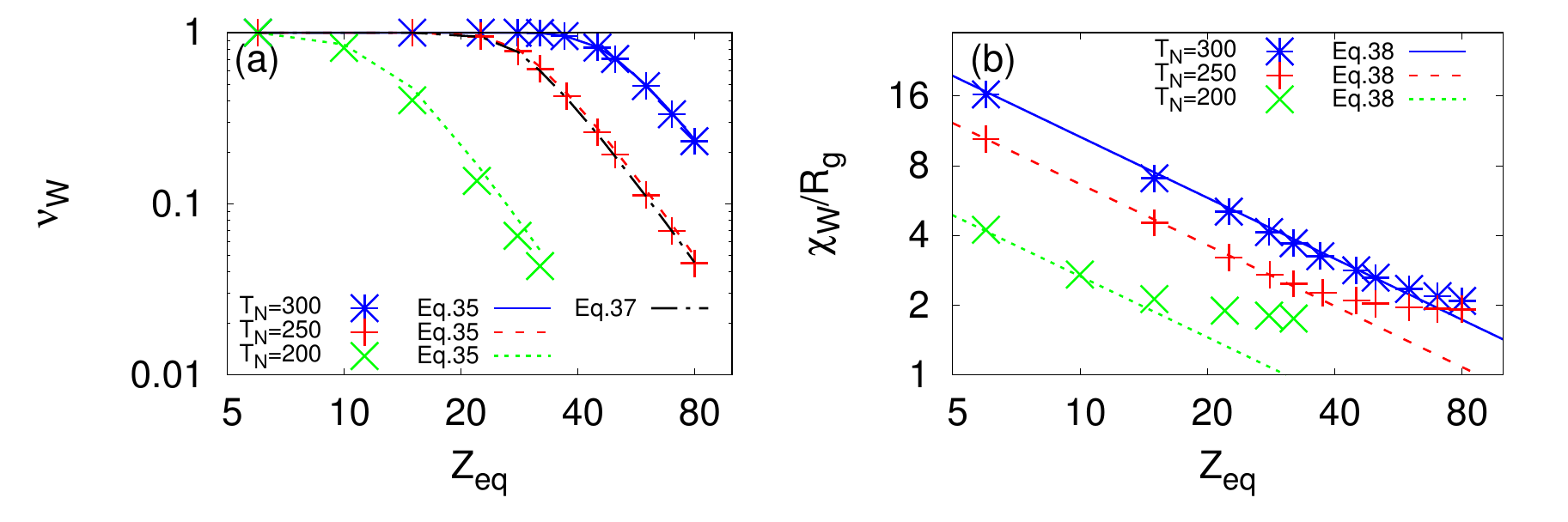}}
\caption{Log-log plot of weld characteristics at $t_g^W$ for a broad range of entanglement numbers $Z_{eq}$, fixed print speed $U_N=10$ mm/s and three print temperatures $T_N =200,250$ and $300^\text{o}$C: (a) weld entanglement $\nu_W$ with lines given Eq. \ref{eq:nuW}. For $T_N=250^\text{o}$C we compare Eq. \ref{eq:nuW} to Eq. \ref{eq:nuW2}. (b) The final weld thickness $\chi_W$ with lines given by Eq. \ref{eq:chiW}. The entanglement number for typical printing materials is in the range $Z_{eq}\sim 20-40$.}
\label{fig:weldZ}
\end{figure*}

For fixed print speed $U_N=10$ mm/s, Figs. \ref{fig:weldZ}a,b show how $\nu_W$ and $\chi_W$ vary with a broader range of molecular weights. Since entanglements only recover once the stretch has relaxed, the final weld entanglement can be predicted from Eq. \ref{eq:nu} as follows:
\begin{subequations}
 \label{eq:integral}
\begin{align}
 \frac{d \nu}{dt} &= \frac{1-\nu}{\tau_d^{eq}(T(t))}, \\
 \int_{\nu_{dep}}^{\nu_W} \frac{1}{1-\nu} d\nu &= \int_{t_w}^{t_g^W} \frac{1}{\tau_d^{eq}(T(t))} dt, 
 \end{align}
\end{subequations}
where $\nu_{dep}$ is the entanglement fraction at the weld site after deposition. Eq. \ref{eq:integral} yields
\begin{equation}
  \nu_W = 1 - (1-\nu_{dep}(Z_{eq}))\exp \left(-  \int_{t_w}^{t_g^W} \frac{1}{\tau_d^{eq}(T(t))} dt \right)
  \label{eq:nuW}
\end{equation}
and is plotted in Fig. \ref{fig:weldZ}a. The integral in Eq. \ref{eq:nuW} can be approximated as
\begin{equation}
  \int_{t_w}^{t_g^W} \frac{1}{\tau_d^{eq}(T(t))} dt \simeq \frac{C}{\tau_d^{eq}(T_N)}\int_{t_w}^{t_g^W} dt,
\end{equation}
where the constant $C$ accounts for $\tau_d^{eq}$ being approximately two orders of magnitude larger when entanglement recovery begins due to cooling (Fig. \ref{fig:relaxstages}b). Thus, the final weld entanglement can be written as
\begin{equation}
 \nu_W \simeq 1 - (1-\nu_{dep}(Z_{eq}))\exp \left( - C \frac{t_g^W}{\tau_d^{eq}(T_N)}  \right),
 \label{eq:nuW2}
\end{equation}
and is also plotted in Fig. \ref{fig:weldZ}a for $T_N =250^\text{o}$C and $C=0.016$. 

$Z_{eq}$ has two effects on entanglement recovery at the weld. First, larger $Z_{eq}$ increases the time scale $\tau_d^{eq}$ so that larger molecular weights diffuse slower and therefore recover fewer entanglements. Second, at a fixed print speed increasing $Z_{eq}$ yields a larger Weissenberg number, which leads to greater disentanglement during deposition i.e. smaller $\nu_{dep}$ \cite{McIlroy:2016}. 



The weld interpenetration depth scales as
\begin{equation}
 \frac{\chi_W}{R_g} \sim \left( \frac{1}{\tau_d^{eq}(T(t))}\right)^{1/4} \sim \frac{1}{Z_{eq}^{7/8}},
 \label{eq:chiW}
\end{equation}
since $\tau_d^{eq}$ scales as $Z_{eq}^{7/2}$ (see Eq. \ref{eq:tdscaling}). The predicted $\chi_W$ deviates from Eq. \ref{eq:chiW}  at some $Z_{eq}$ depending on print temperature (Fig. \ref{fig:weldZ}b). This critical $Z_{eq}$ defines the molecular weight at which diffusion is arrested during relaxation of the tube stretch. In this regime the tube dynamics reduce to the reptation time via Eq. \ref{eq:taud} and speed up diffusion. This effect results in a larger $\chi_W$ than predicted by Eq. \ref{eq:chiW}. 

\subsection{Weld Fracture Toughness}

Mechanical strength at the weld is determined in part by the molecular weight between entanglements (Eq. \ref{eq:GeGc}). In the weld region we have
\begin{equation}
 M_e^W = \frac{M_e}{\nu_W}. 
 \label{eq:MeW}
\end{equation}
Thus, from Eq. \ref{eq:GeGc}b the fracture toughness at the weld is given by
\begin{equation}
 G_c^W \sim \left(1- \frac{M_e^W}{q M_w}\right)^2 \sim \left(1 - \frac{1}{q\nu_W Z_{eq}} \right)^2.
 \label{eq:GcW}
\end{equation}
Hence decreasing $\nu_W$ increases $M_e^W$ and reduces the toughness of the weld. The equilibrium bulk strength $G_c$ is only achieved for $\nu_W=1$. 

\begin{figure}[t]
 \centerline{\includegraphics[width=9cm]{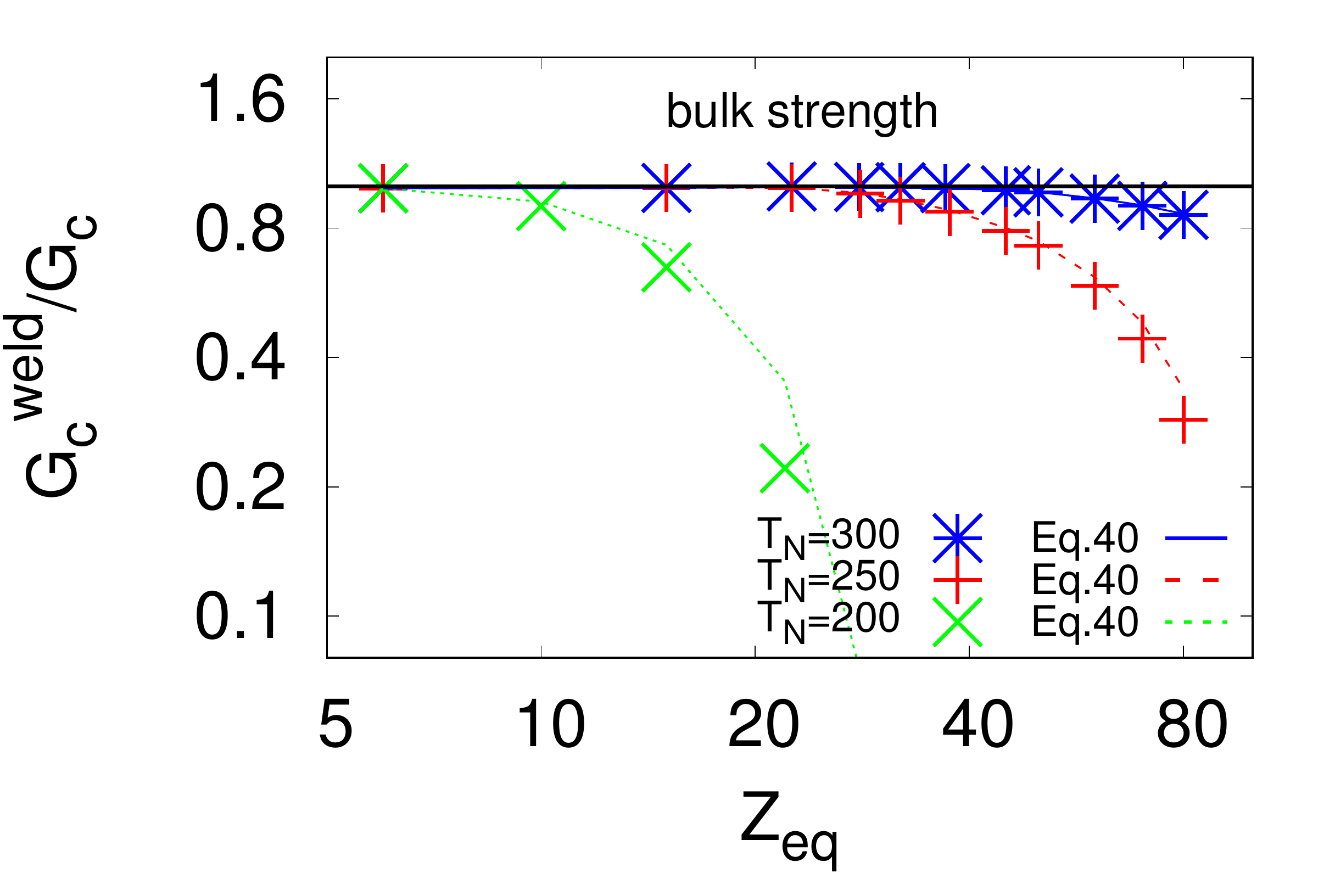}}
\caption{Log-log plot of the fracture toughness at the weld $G_c^W$ normalised by the equilibrium fracture toughness $G_c$ for fixed print speed $U_N=10$ mm/s and three print temperatures $T_N =200,250$ and $300^\text{o}$C. The entanglement number for typical printing materials is in the range $Z_{eq}\sim 20-40$. Lines are given by Eq. \ref{eq:GcW}. }
\label{fig:weldMe}
\end{figure}

For polycarbonate, Fig. \ref{fig:weldMe} shows that $G_c^W$ can decrease by 50\% depending on the entanglement number $Z_{eq}$ and the print temperature $T_N$. In particular, for a prescribed print temperature, there exists a maximum molecular weight that can be printed whilst maintaining $G_c^W = G_c$. For example, $Z_{eq} \le 40$ can be printed at $T_N=300^\text{o}$C before strength is affected by disentanglement, whereas only $Z_{eq} \le 22$ maintains bulk strength at lower temperature $T_N=200^\text{o}$C.



\section{Conclusion}

We have developed a model for the non-isothermal FFF welding process to test the effect of changing print speed, temperature and entanglement number on the ultimate welding characteristics of an non-crystalline polymer melt. It has previously been shown that the extrusion process can significantly deform and disentangle the polymer microstructure prior to welding \cite{McIlroy:2016}. After deposition the printed layer cools and this deformation relaxes via reptation, whilst inter-diffusing with the previously-printed layer. The temperature profile at the weld between the two layers is calculated by solving the heat equation in one dimension. The cooling rate inhibits the total relaxation of the deformation induced by printing, so that the ultimate structure of the weld is anisotropic and less entangled than the equilibrium material for typical printing conditions. Solving a diffusion equation that incorporates anisotropic and mutual diffusion yields the thickness of the weld formed between the layers. 

The model predicts that the weld thickness typically surpasses $R_g$, but not quite enough to fully relax. Thus, mechanical strength should not be limited by interpenetration depth. However, despite sufficient weld thickness for bulk strength at the interface, entanglements do not have sufficient time to recover during cooling; $\nu_W$ is as low as $50$\% for typical printing conditions. Since a disentangled weld structure can significantly increase the entanglement molecular weight in the weld region, the mechanical properties at the weld may be significantly reduced. These findings suggest that disentanglement in the nozzle combined with the delay in entanglement recovery due to the relaxation of the tube stretch is the key mechanism responsible for a reduced mechanical strength in the weld region.  


Although the theory of flow-induced disentanglement has been compared to molecular dynamics simulations of planar shear flow \cite{Ianniruberto:2015}, the way in which polymers recover entanglements in the event of flow cessation is yet to be addressed. It is crucial to benchmark this re-entanglement theory against molecular dynamics simulations of disentangled melts, in particular to verify that $\nu<1$ for $\chi> R_E$ is a result of the initial delay in entanglement recovery due to tube stretch. 

Practically, thicker and more entangled welds can be formed by increasing the print temperature or using a less entangled printing material, since both parameters significantly reduce the reptation time. 
We find that the weld thickness is independent of the deformation induced by different nozzle shear rates, as is the final entanglement network for $Z_{eq}=22$. Consequently equivalent mechanical integrity is expected across all print speeds. Thus, the maximum print speed available can be exploited to increase productivity.


\section*{Acknowledgements}

We thank Jonathan Seppala and Kalman Migler for advice and an enjoyable collaboration, as well as the National Institute for Standards and Technology (NIST), Georgetown University, and the Ives Foundation for funding.

\appendix

\setcounter{figure}{0}

\section{Extrusion and Deposition Model}
\label{sec:appendix_model}

\begin{table*}[t]
\caption{Material parameters for a typical amorphous printing material, polycarbonate.}
\centerline{\begin{tabular}{l|c|c|c}
\hline
\hline
 {\bf Polycarbonate Properties} & Notation & Value & Units \\
\hline
\hline
{Glass-Transition Temperature} & $T_g$ & 140 & $^\text{o}$C \\
{Thermal Diffusivity \cite{Zhang:2002}} (at 25$^\text{o}$C) & $\alpha$ & 0.14 & mm$^2$/s \\
Eq. \ref{eq:alpha}  & $\alpha_0$ & 0.16 &mm$^2$/s \\
Eq. \ref{eq:alpha} & $B$ & $1.3 \times 10^{-3}$ & $^\text{o}\text{C}^{-1}$\\
\hline
{Molecular Weight} & $M_w$ & 60 & kDa \\
{Entanglement Molecular Weight \cite{Mark:1996}} & $M_e$ & 1.6 & kDa \\
{Plateau Modulus \cite{Mark:1996} } & $G_e$ & $2.6 \times 10^6$ & Pa \\
{Entanglement Time \cite{DoiEdwards:1986} (at $T_0$)} &$ \tau_e^0$ & $3.3 \times 10^{-7}$  & s \\
\hline
{Reference Temperature} & $T_0$ & 260& $^\text{o}$C \\
{WLF parameter } & $C_1$ & 3  & - \\
{WLF parameter } & $C_2$ & 160 & - \\
\hline
{Equilibrium Entanglement Number} & $Z_{eq}$ & 37 & - \\
{CCR parameter } & $\beta$ & 0.3 & - \\
\hline
{Equilibrium Reptation Time} (at $T_N$) & $ \tau_d^{eq}$ & 0.033 & s \\
{Equilibrium Rouse Time} (at $T_N$) & $\tau_R^{eq}$ & 0.00057 & s \\
\hline
\end{tabular}}
\label{tab:polycarbonate}
\end{table*}

\begin{table*}[t!]
\caption{Model printing parameters for two typical print speeds; a `fast' and a `slow' case.}
\centerline{\begin{tabular}{l|c|c|c|c}
\hline
\hline
 {\bf Printing Parameters} & Notation & Fast Case & Slow Case & Units \\
\hline
\hline
{Extrusion Temperature} & $T_N$     & 250 &  250 & $^\text{o}$C \\
{Ambient Temperature} & $T_a$     & 95 & 95 &  $^\text{o}$C \\
\hline
{Mean Extrusion Speed} & $U_N$                          & 0.075 & 0.01      &  m/s     \\
{Mean Print Speed} & $U_L$                              & 0.100 & 0.013   &  m/s  \\
\hline
{Nozzle Radius}  & $R$ & 0.2 & 0.2 & mm \\
{Layer Thickness} & $H$ & 0.3 & 0.3 & mm \\
\hline
{Reptation Weissenberg Number (average)} & $\overline{Wi}$ & 13 & 2 & -\\
{Rouse Weissenberg Number (average)} & $\overline{Wi}_R$ &     0.07 & 0.0009 & - \\
\hline
\end{tabular}}
\label{tab:speeds}
\end{table*}

Here we summarise the FFF extrusion and deposition model detailed in Ref. \cite{McIlroy:2016}.

The printing material is heated to temperature $T_N$ and extruded through a nozzle of radius $R$ at mass-averaged speed $U_N$. Assuming a steady state the momentum balance is given by
\begin{equation}
 \nabla \cdot \boldsymbol{\sigma} =0,
 \label{eq:momentum}
\end{equation}
for stress tensor $\boldsymbol{\sigma}$. The total stress in the polymer melt comprises solvent and polymer contributions
\begin{equation}
\boldsymbol{\sigma} = - p {\bf I} + G_e ({\bf A} - {\bf I}) + 2 \mu_s ( {\bf K} + {\bf K}^T),
\label{eq:stress}
\end{equation}
where $p$ is the isotropic pressure and $G_e$ is the plateau modulus. For times shorter than $\tau_e$, Rouse modes corresponding to lengths shorter than $M_e$ contribute to a background viscosity defined as \citep{Graham:thesis}
\begin{equation}
 \mu_s = \frac{\pi^2}{12} \frac{G_e}{Z_{eq}} \tau_R^{eq}.
\end{equation}

The temperature profile is assumed to be uniform across the nozzle and Eq. \ref{eq:momentum} is solved alongside the Rolie-Poly Eqs. (\ref{eq:Rolie-Poly}) and (\ref{eq:nu}) to calculate the plug-like velocity profile, and the polymer deformation and disentanglement across the nozzle. Nozzle Weissenberg numbers (see Eqs. (\ref{eq:WiN}) and (\ref{eq:WiNR})) for typical fast and slow print speeds are quoted in Table \ref{tab:speeds}; the polymer is found to stretch and orient in the nozzle depending on print speed.  

The material is then deposited into a layer of thickness $H$, which travels horizontally at speed $U_L$, in a frame moving with the nozzle. During this deposition the material must speed up and deform to make a $90^\text{o}$ turn and transform from circular to elliptical geometry (since typical $H < 2R$). In order to conserve mass, $U_L > U_N$ (Eq. \ref{eq:massconservation}). 


Rather than calculating the full fluid mechanics, the shape of the deposition is prescribed (ignoring die swell) and the fluid is advected using local flux conservation. We assume a uniform temperature profile through out the deposition and neglect polymer relaxation during this stage. Due to the assumption of zero polymer relaxation, the model breaks down for slower print speeds such that $\overline{Wi} < H/R$ \cite{McIlroy:2016}. In this way the velocity profile ${\bf u}$ is prescribed only by the geometry of the deposition and the Rolie-Poly Eq. \ref{eq:Rolie-Poly} reduces to
\begin{equation}
 ({\bf u} \cdot \nabla ) {\bf A} = {\bf K} \cdot {\bf A} + {\bf A} + {\bf K}^T.
\end{equation}

\section{Weld Structure and Print Speed}
\label{sec:appendix_speed}

\begin{figure*}[p!]
\centerline{\includegraphics[width=12cm]{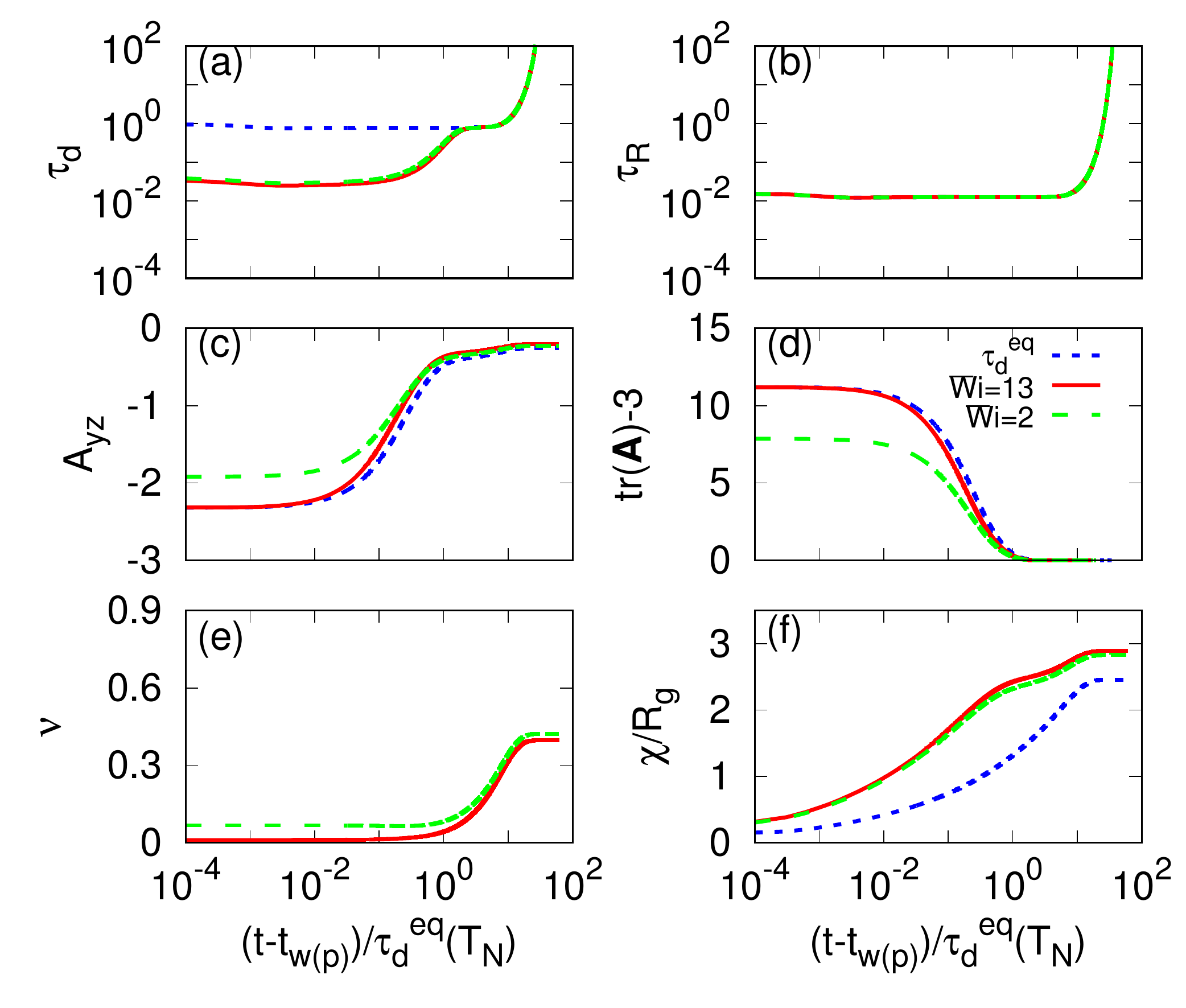}}
\caption{Relaxation dynamics at weld site $\mathtt{b_p}$ for $\overline{Wi}=2$ and 13: (a) the reptation time $\tau_d$ (Eq. \ref{eq:taud}), (b) the Rouse time $\tau_R$ (Eq. \ref{eq:tauReq}), (c) the principle shear component $A_{yz}$ and (d) the tube stretch $\text{tr}{\bf A}-3$, (e) entanglement fraction $\nu$ and (f) the interpenetration distance $\chi$ (Eq. \ref{eq:weldthickness}). Parameters: $T_N=250^\text{o}$C, $Z_{eq}=37$ and $\overline{Wi}=2$.}
\label{fig:relax}
\centerline{\includegraphics[width=12cm]{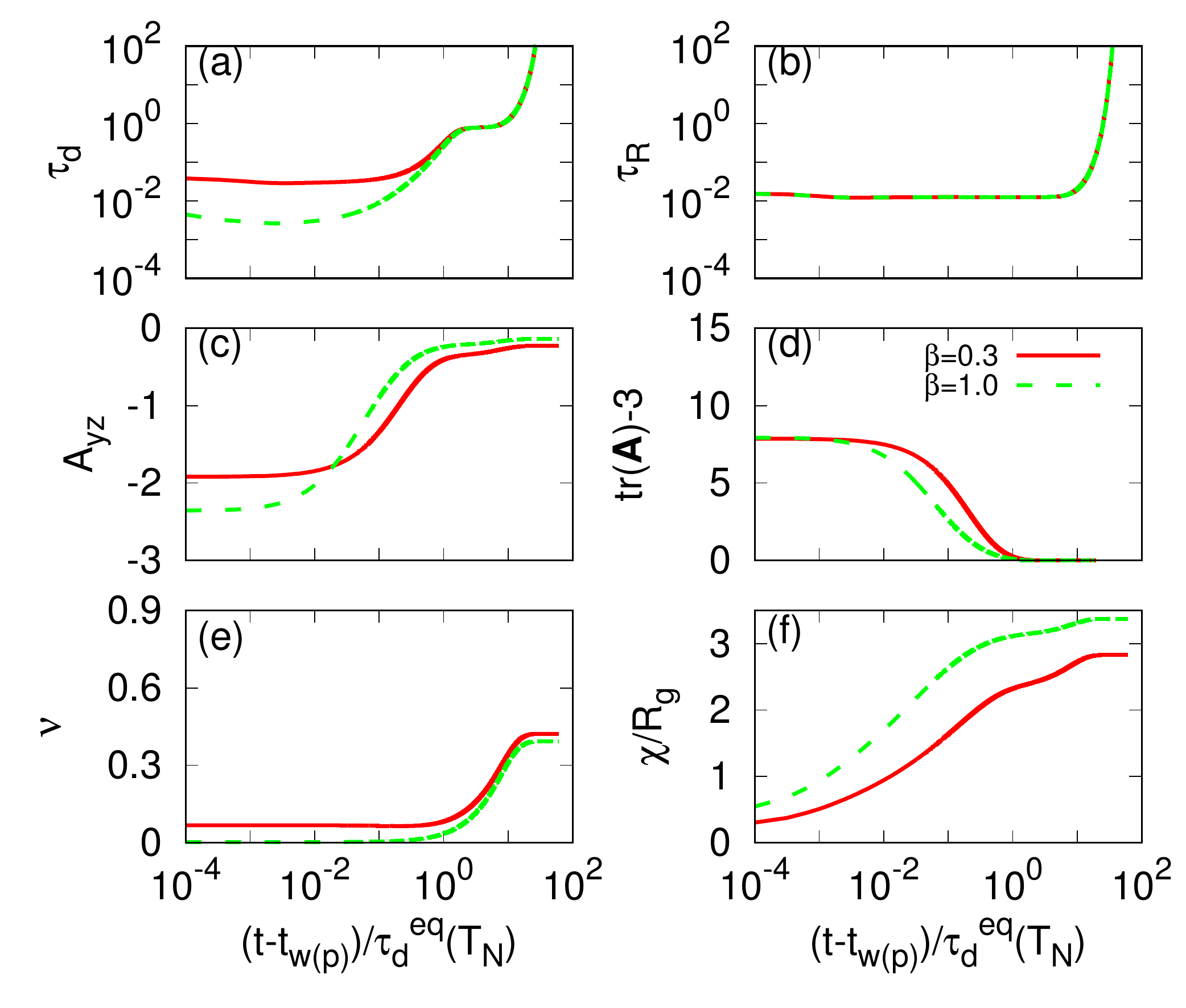}}
\caption{Relaxation dynamics at weld site $\mathtt{b_p}$ for $\beta=0.3$ and 1: (a) the reptation time $\tau_d$ (Eq. \ref{eq:taud}), (b) the Rouse time $\tau_R$ (Eq. \ref{eq:tauReq}), (c) the principle shear component $A_{yz}$ and (d) the tube stretch $\text{tr}{\bf A}-3$, (e) entanglement fraction $\nu$ and (f) the interpenetration distance $\chi$ (Eq. \ref{eq:weldthickness}). Parameters: $T_N=250^\text{o}$C, $Z_{eq}=37$ and $\overline{Wi}=2$.}
\label{fig:relaxbeta}
\end{figure*}

Fig. \ref{fig:relax} shows the relaxation dynamics at the weld site $\mathtt{b_p}$ for the two typical print speeds given in Table \ref{tab:speeds}, corresponding to $\overline{Wi}=2$ and 13. Parameters are $T_N=250^\text{o}$, $Z_{eq}=37$ and $\beta=0.3$. Time is scaled by $\tau_d^{eq}(T_N)$ in order to highlight the effect of the cooling temperature profile and the case $\tau_d = \tau_d^{eq}$ (with an initial polymer configuration imposed by the fast print speed) is plotted to highlight the effect that polymer stretch has on the relaxation process. 

The reptation time $\tau_d(T,\dot{\gamma})$ and the Rouse time $\tau_R(T)$ are plotted in Figs. \ref{fig:relax}a,b. Although the stretch induced by printing significantly reduces the reptation time compared to $\tau_d^{eq}$ (Eq. \ref{eq:taud}), the difference in the reptation for the two typical print speeds is small. The Rouse time is independent of print speed (Eq. \ref{eq:tauReq}). 

Figs. \ref{fig:relax}c,d  shows the relaxation of the principle shear component $A_{yz}$ and the the tube stretch $\text{tr}{\bf A}-3$. The principle shear component $A_{yz}$ relaxes on a time scale $t/\tau_d^{eq}(T_N)>1$, demonstrating how the cooling temperature profile inhibits the relaxation process. We see a similar two-mode relaxation (Rouse followed by reptation) for both print speeds. The stretch has sufficient time to relax prior to the glass transition but the polymer orientation remains out of equilibrium for both print speeds. The structure is slightly closer to equilibrium for the fast-printing case due to a slightly smaller reptation time at $t_{w(p)}$. 

Fig. \ref{fig:relax}e shows how the entanglement fraction evolves at the weld for the two print cases. The weld region is approximately 50\% less entangled than the bulk material and the final $\nu$ is similar for both printing speeds. Finally, Fig. \ref{fig:relax}f shows the isotropic interpenetration distance $\chi$ calculated by Eq. (\ref{eq:weldthickness}). Since the reptation time is initially smaller, faster printing allows for a slightly longer welding time and consequently a slightly thicker weld form, although the difference is much smaller than the polymer size ($\ll R_g$).

\section{Effect of CCR parameter}
\label{sec:appendix_CCR}

Fig. \ref{fig:relaxbeta} shows the relaxation dynamics at the weld site $\mathtt{b_p}$ for CCR parameters $\beta=0.3$ and 1. Parameters are $T_N=250^\text{o}$C, $Z_{eq}=37$ and $\overline{Wi}=2$. The reptation time (Fig. \ref{fig:relaxbeta}a) depends on $\beta$ whilst the tube is stretched ($\text{tr}{\bf A}>3$), and $\beta=1$ reduces the reptation time by an order of magnitude compared to $\beta=0.3$. Fig \ref{fig:relaxbeta}b shows that the Rouse time is independent of $\beta$ (Eq. \ref{eq:tauReq}). Due to the reduced reptation time, both the principle shear deformation $A_{yz}$ and the tube stretch $\text{tr}{\bf A}$ relax faster for $\beta=1$ (Fig. \ref{fig:relaxbeta}c,d). Since entanglement recovery depends only on $\tau_d^{eq}(T)$, fewer entanglements are recovered for $\beta=1$ due to the smaller initial $\nu$ generated (Fig. \ref{fig:relaxbeta}e). The polymer is able to diffuse further in the case $\beta=1$, although the difference is less than $R_g$ (Fig. \ref{fig:relaxbeta}f), again as a result of the reduced reptation time. The conclusion that weld properties are independent of print speeds also holds for $\beta=1$.

\bibliographystyle{elsarticle-num-names}
\bibliography{references}

\end{document}